\documentclass[]{aa} 
\usepackage{graphicx}
\usepackage{xcolor}
\usepackage{natbib}  
\usepackage{scalefnt}   
\usepackage[varg]{txfonts}
\usepackage{hyperref}

\newcommand{\Teff}{\ensuremath{T_\mathrm{eff}}}
\newcommand{\gf}{\ensuremath{gf}}

\newcommand{\logg}{\ensuremath{\log\,g}}
\def\teff{$T\rm_{eff}$}
\newcommand{\kms}{$\rm km\,s ^{-1}$}
\def\ss  {$\sigma$}
\def\Feu  {\ion{Fe}{i}} 
\def\Fed  {\ion{Fe}{ii}}
\def\Cou  {\ion{Co}{i}}
\def\Bad  {\ion{Ba}{ii}}
\def\vt   {$v_{t}$}
\begin{document} 

\title{The metal-poor end of the Spite plateau: II Chemical and dynamical investigation
\thanks{Based on observations collected at the European Organisation for Astronomical Research in the Southern Hemisphere (Programmes 076.A-0463 PI(Lopez), 077.D-0299 PI(Bonifacio)), 086.D-0871(A) (PI Mel\'endez).}
\thanks{The on-line table with equivalent widths discussed in this paper is only available in electronic form
at the CDS via anonymous ftp to cdsarc.u-strasbg.fr (130.79.128.5)
or via \url{http://cdsweb.u-strasbg.fr/cgi-bin/qcat?J/A+A/}}
}

\author{
A.M.~Matas Pinto \inst{1}\and
M.~Spite \inst{1} \and
E.~Caffau    \inst{1} \and
P.~Bonifacio \inst{1} \and
L.~Sbordone \inst{2} \and
T.~Sivarani \inst{3} \and
M.~Steffen \inst{4} \and
F.~Spite \inst{1} \and
P.~Fran\c{c}ois \inst{5,6} \and
P.~Di Matteo  \inst{1} 
}
\institute{GEPI, Observatoire de Paris, Universit\'{e} PSL, CNRS,  5 Place Jules Janssen, 92190 Meudon, France
\and
ESO - European Southern Observatory, Alonso de Cordova 3107, Vitacura, Santiago, Chile
\and
Indian Institute of Astrophysics, India
\and
Leibniz-Institut f\"{u}r Astrophysik Potsdam, D-14482 Potsdam, Germany
\and
GEPI, Observatoire de Paris, Universit\'{e} PSL, CNRS,  77 Av. Denfert-Rochereau, 75014 Paris, France
\and
UPJV, Universit\'e de Picardie Jules Verne, 33 rue St Leu, 80080 Amiens, France}

\date{Received May 11, 2021; accepted August 22, 2021}

\abstract
{The study of old, metal-poor stars deepens our knowledge on the early stages of the universe. In particular, the study of these stars gives us a valuable insight into the masses of the first massive stars and their emission of ionising photons.}
{We present a detailed chemical analysis and determination of the kinematic and orbital properties of a sample of 11 dwarf stars. These are metal-poor stars, and a few of them present a low lithium content. We inspected whether the other elements also present anomalies.}
{We analysed the high-resolution UVES spectra of a few metal-poor stars using the Turbospectrum code to synthesise spectral lines profiles. This allowed us to derive a detailed chemical analysis of Fe, C, Li, Na, Mg, Al, Si, CaI, CaII, ScII, TiII, Cr, Mn, Co, Ni, Sr, and Ba. }
{We find excellent coherence with the reference metal-poor First Stars sample. The lithium-poor stars do not present any anomaly of the abundance of the elements other than lithium.
Among the Li-poor stars, we show that  CS\,22882-027  is very probably a blue-straggler. The star CS\,30302-145, which has a Li abundance compatible with the plateau, has a very low Si abundance and a high Mn abundance. In many aspects, it is similar to the $\alpha$-poor star HE\,1424-0241, but it is less extreme. It could have been formed in a satellite galaxy and later been accreted by our Galaxy. This hypothesis is also supported by its kinematics.}
{}
   \maketitle
%
\section{Introduction}

The lithium observed in metal-poor stars is built up during the Big Bang nucleosynthesis \citep[see e.g.][]{pitrouCU18}. It is a very fragile element that is destroyed by proton fusion when the temperature reaches about $2~10^{6}$\,K. In a main-sequence star, this element is destroyed when the convective zone of the atmosphere reaches the layers in which the temperature is higher than its characteristic fusion temperature. However, in the atmosphere of warm metal-poor dwarf stars, the convective zone is not so deep, and the initial lithium abundance is expected to be preserved. We  refer to standard Big Bang nucleosynthesis (SBBN) 
when we assume that the Universe at the  time that primordial nucleosynthesis occurred was homogeneous and isotropic,
and there were only three light neutrinos. Under these assumptions, the
amount of lithium produced is a non-monotonic function of baryonic density, or equivalently, of the baryon-to-photon ratio. The Planck measurements of the
baryonic density \citep{PlanckColl16} imply that the lithium abundance in the Universe after the Big Bang was $ \rm A(Li)=2.75 \pm 0.02$ dex \citep{pitrouCU18}. In warm dwarf stars in the metallicity range $ \rm -2.9<[Fe/H]<-1.5 $, the observed value of the lithium abundance is constant \citep[the so-called Spite plateau, as noted by][]{spite82,spite82a} and close to A(Li)=2.1\,dex \citep{bonifacio07}. This value is over three times lower than the expected primordial value.

\citet{bonifacio07} studied the lithium abundance of a sample of 18 very and extremely metal-poor (VMP and EMP) turnoff stars ($ \rm -3.5 < [Fe/H]<-2.5$). All the stars with a metallicity higher than [Fe/H]=--2.9 had a lithium abundance compatible with the Spite plateau within the
measurement uncertainties, but surprisingly, in the 10 stars with ($[{\rm Fe/H}]<-3.0,$ the scatter of the Li abundance was large. The Li abundance was sometimes at the level of the plateau, but sometimes significantly below, never above. It was then decided to increase the sample of the metal-poor turnoff stars in order to refute or confirm the behaviour of the lithium abundance at low metallicity. A new sample of 11 metal-poor stars was then studied \citep[][hereafter Paper I]{sbordone10}, the abundances of iron and lithium were computed in this sample, and a meltdown (see Paper I)  of the lithium plateau at low metallicity was confirmed \citep[see also e.g.][]{Aoki09,Matsuno17,AguadoGA19}.\\
The disagreement between the observed abundance of Li in the stars of the plateau and the abundance predicted by the standard Big Bang is known as  the cosmological lithium problem and has recently been addressed by \citet{MolaroCF20} and \citet{SimpsonMB20}, for instance.  

The aim of this new paper is to study the other problem that might be related to this: the variable decrease in Li abundance at metallicities lower than ${\rm [Fe/H]}\sim -3$.
We present here a complete analysis of the abundance of the elements from Li to Ba in the 1 VMP or EMP stars studied in Paper I, and search for any definite correlation between the lithium abundance and the abundance of the other elements in these stars, which could constrain the behaviour of the lithium abundance at extremely low metallicity. We also investigate whether there is any correlation between the kinematic  properties of the stars and their Li abundance.

Many of the stars of this sample have already been studied (see Appendix A), but it was important to have a homogeneous study of all the stars of this sample. In this appendix we only report recent analyses based on high-resolution spectra.

\section{Observations} 

The observations of the stars we studied are described in detail in Paper I (see their Table\,1). Briefly, observations were performed with the high-resolution spectrograph UVES \citep{Dekker00} at the ESO-VLT. The spectra have a resolving power $R \simeq 40\,000$ and were centred at 390\,nm  (spectral range: 330 - 451\,nm) and 580\,nm (spectral range:  479 - 680 nm). For two stars (BS\,17572-100 and HE\,1413-1954) that were previously studied in the frame of the HERES program \citep{Christlieb04,barklem05} from UVES spectra centred at 437nm (spectral range: 376 - 497\,nm), the blue spectra were centred at 346\,nm (spectral range: 320 - 386\,nm). The S/N of the spectra at 400\,nm is only about half of the S/N measured at 670\,nm (see Table\,1 in Paper I) and thus generally does not exceed 50. 
For two stars, CS 22188-033 and HE\,0148-2611, new UVES spectra from the ESO archives, centred at 390 and
 580\,nm, were also used, increasing the S/N ratio of the mean spectrum. The data were reduced using the standard UVES pipeline with the same procedures as used in \citet{bonifacio07}.

\section{Analysis}
\subsection{Stellar parameters}   \label{sec:stelpar}

\begin{table*}
\centering 
\scalefont{0.75}
\caption{Photometric data of the sample, parameters of the atmosphere models, and resulting \Feu~ and \Fed~ abundances.}
\label{tab:ParametrosEstelares}
\begin{tabular}{ccccccccccccccc}
\hline
\hline\\
              &            &  paral.   &  paral. $/$ &   d    &   g     &        &         &         &              & \Teff &  \logg  & \vt   &[Fe1/H]&[Fe2/H]\\
Star          &  parallax  &  error    & paral err   &  (pc)  &   mag   & (BP-RP)&  A(V)   & G$_{0}$ &(BP-RP)$_{0}$ &   K   &  cgs    & km$/$s& \\
\hline\\
BS\,17572-100 &  2.184     &  0.046    &  0.02       &  458   &   12.14 &  0.619 & 0.11554 &  3.80   &    0.57      & 6450  &  4.15   & 1.5&-2.61&-2.50   \\
CS\,22188-033 &  2.236     &  0.039    &  0.02       &  447   &   13.07 &  0.649 & 0.04077 &  4.80   &    0.63      & 6230  &  4.45   & 1.4&-3.03&-2.96   \\
CS\,22882-027 &  0.754     &  0.057    &  0.08       &  1327  &   15.04 &  0.552 &     -   &  4.39   &    0.54      & 6630  &  4.45   & 1.4&-2.55&-2.41   \\
CS\,22950-173 &  1.300     &  0.047    &  0.04       &  770   &   13.91 &  0.666 & 0.14108 &  4.44   &    0.61      & 6320  &  4.30   & 1.4&-2.88&-2.54   \\
CS\,29491-084 &  0.895     &  0.036    &  0.04       &  1117  &   13.37 &  0.611 & 0.04238 &  3.12   &    0.59      & 6340  &  3.90   & 1.8&-2.97&-2.88   \\
CS\,29514-007 &  0.845     &  0.029    &  0.03       &  1184  &   13.83 &  0.631 & 0.07363 &  3.45   &    0.60      & 6320  &  4.00   & 1.7&-2.88&-2.73   \\
CS\,29516-028 &  1.311     &  0.057    &  0.04       &  763   &   14.77 &  0.876 & 0.39730 &  5.25   &    0.71      & 5960  &  4.50   & 1.2&-3.44&-3.26   \\
CS\,30302-145 &  0.844     &  0.041    &  0.05       &  1185  &   14.34 &  0.633 & 0.16563 &  3.93   &    0.56      & 6480  &  4.20   & 1.7&-3.06&-2.87   \\
CS\,30344-070 &  0.691     &  0.026    &  0.04       &  1447  &   14.34 &  0.570 & 0.04046 &  3.52   &    0.55      & 6510  &  4.05   & 1.8&-3.01&-2.82   \\
HE\,0148-2611 &  0.822     &  0.024    &  0.03       &  1216  &   14.35 &  0.574 & 0.04222 &  3.91   &    0.56      & 6510  &  4.16   & 1.5&   \\
HE\,1413-1954 &  0.542     &  0.065    &  0.12       &  1846  &   15.18 &  0.671 & 0.26911 &  3.78   &    0.56      & 6480  &  4.15   & 1.9&   \\
\hline
\multicolumn{15}{l}{The distance of the stars was computed by a simple inversion of the parallax.}\\
\end{tabular}
\end{table*} 


\begin{table*}
\caption{Abundance of the elements in the Sun and in our sample of stars.}
\label{tab:Abundance}
\centering
\scalefont{0.67}
\begin{tabular}{l@{~~~~~~}c@{~~~~~~}c@{~~~~~~}c@{~}r@{~~~~~~~~}c@{~}c@{~}r@{~~~~~~~}c@{~}r@{~~~~~~~~}c@{~}r@{~~~~~~~~}c@{~}c@{~}r@{~~~~~~~}c@{~}r@{~~~~~~~~}c@{~}c@{~}r@{~~~~~~~}c@{~}r@{~}}
\hline

Star           & [Fe/H]    &  A(C) & A(Na)   & [Na/Fe]  & A(Mg)  &\ss~~~N&[Mg/Fe]   & A(Al)& [Al/Fe] & A(Si) &  [Si/Fe]& A(Ca1)& \ss~~N& [Ca1/Fe]& A(Ca2)&[Ca2/Fe]&A(Sc2)&[Sc2/Fe]\\
\hline 
     Sun       &          &   8.5   & 6.30   &          & 7.54   &       &        &  6.47 &         & 7.52  &         & 6.33  &        &       &        &      &  3.10  &      \\       
BS\,17572-100  & --2.60   & $<6.23$ & 3.42   &  --0.29  & 5.26   & 0.12~8&   0.31 & 3.75  & --0.13  & 4.69  & --0.24  & 4.27  & 0.06~7 &  0.54 &  4.25  & 0.52 &  0.74  &  0.24\\ 
CS\,22188-033  & --2.99   & $<6.22$ & 2.98   &  --0.34  & 4.87   & 0.07~7&   0.31 & 3.48  & --0.00  & 4.85  &   0.32  & 3.86  & 0.05~4 &  0.52 &  3.91  &  -   &  0.29  &  0.17\\ 
CS\,22882-027  & --2.49   & $<6.20$ & 3.17   &  --0.43  & 5.19   & 0.12~6&   0.14 & 3.78  & --0.21  & 5.04  &   0.00  & 4.33  & 0.11~5 &  0.49 &  4.25  & 0.40 &  0.32  &--0.30\\ 
CS\,22950-173  & --2.71   & $<6.02$ & 3.01   &  --0.56  & 5.02   & 0.06~6&   0.18 & 3.59  & --0.18  & 4.79  & --0.03  & 3.94  & 0.03~4 &  0.31 &  3.99  &  -   &  0.44  &  0.04\\ 
CS\,29491-084  & --2.93   & $<6.05$ & 3.10   &  --0.22  & 4.80   & 0.07~6&   0.19 & 3.45  & --0.09  & 4.80  &   0.21  & 4.12  & 0.20~6 &  0.72 &  4.07  & 0.67 &  0.49  &  0.32\\ 
CS\,29514-007  & --2.81   & $<6.33$ & 3.05   &  --0.39  & 4.97   & 0.10~6&   0.23 & 3.53  & --0.13  & 4.84  &   0.13  & 4.04  & 0.08~6 &  0.52 &  4.17  &  -   &  0.62  &  0.33\\ 
CS\,29516-028  & --3.35   & $<6.00$ & 2.60   &  --0.35  & 4.43   & 0.04~5&   0.24 & 3.14  &   0.02  & 4.30  &   0.13  & 3.42  & 0.16~3 &  0.44 &  3.33  & 0.35 &  0.03  &  0.28\\ 
CS\,30302-145  & --2.96   & $<6.15$ & 2.53   &  --0.90  & 4.37   & 0.09~5& --0.21 & 2.84  & --0.67  & 4.07  & --0.49  & 3.82  & 0.04~2 &  0.45 &  3.57  & 0.20 &  0.32  &  0.18\\ 
CS\,30344-070  & --2.92   & $<6.51$ & 3.01   &  --0.38  & 4.75   & 0.13~6&   0.12 & 3.49  & --0.06  & 4.49  & --0.12  & 4.02  & 0.19~3 &  0.61 &  4.08  & 0.66 &  0.38  &  0.19\\ 
HE\,0148-2611  & --3.21   & $<6.45$ & 2.39   &  --0.65  & 4.30   & 0.06~6& --0.04 & 2.97  & --0.30  & 4.17  & --0.15  & 3.67  & 0.16~4 &  0.55 &  3.53  & 0.41 &  0.00  &  0.11\\ 
HE\,1413-1954  & --3.29   & $ 6.9 $ & 2.72   &  --0.29  & 4.67   & 0.22~5&   0.41 & 3.05  & --0.14  & 4.01  & --0.23  & 3.38  & 0.10~3 &  0.34 &  3.67  & 0.62 &  0.07  &  0.25\\
\\
\end{tabular}
\begin{tabular}{l@{~~~~~~~}c@{~~}c@{~~}r@{~~~~~~~~}c@{~~}c@{~~}r@{~~~~~~~~}c@{~~}r@{~~~~~~~~}c@{~~}c@{~~}r@{~~~~~~~~}c@{~~}c@{~~}r@{~~~~~~~~}c@{~~}r@{~~~~~~~~}c@{~~}r@{~~~~}c@{~~~~}}
\hline 
Star     & A(Ti2) & \ss~~~N  &[Ti2/Fe]& A(Cr)  &\ss~~~N & [Cr/Fe]& A(Mn)& [Mn/Fe] & A(Co)& \ss~~~N&[Co/Fe]& A(Ni)&\ss~~~N&[Ni/Fe] & A(Sr)  &  [Sr/Fe] & A(Ba) & [Ba/Fe]\\
\hline 
   Sun        & 4.90  &          &       &  5.64  &        &       &  5.37 &         &  4.92&       &        & 6.23&        &         & 2.92   &          & 2.17   &        \\            
BS\,17572-100 &  2.80 & 0.09~23  &   0.50&  3.04  & 0.16~7 & --0.01& 2.12  &  --0.66 & 2.69 & 0.01~2&  0.37  & 3.73&~~-~~~~1&   0.09  &   0.27 &  --0.06  & --0.91 &  --0.49\\  
CS\,22188-033 &  2.29 & 0.08~10  &   0.37&  2.44  & 0.07~5 & --0.22& 1.73  &  --0.66 & 2.23 & 0.04~3&  0.29  & 3.30& 0.15~3 &   0.05  & --0.39 &  --0.32  & --1.22 &  --0.41\\  
CS\,22882-027 &  2.83 & 0.09~12  &   0.41&  3.01  & 0.06~5 & --0.15& 2.32  &  --0.57 & 2.86 & 0.19~3&  0.42  & 3.80&~~-~~~~1&   0.05  & --1.18 &  --1.62  &   --   &    --  \\  
CS\,22950-173 &  2.41 & 0.10~10  &   0.21&  2.61  & 0.06~5 & --0.33& 1.86  &  --0.81 & 2.37 & 0.16~3&  0.15  & 3.47&  0.24~3& --0.06  & --0.52 &  --0.74  & --1.39 &  --0.86\\  
CS\,29491-084 &  2.55 & 0.04~15  &   0.58&  2.61  & 0.05~5 & --0.10& 1.83  &  --0.61 & 2.39 & 0.18~2&  0.40  & 3.38&  0.17~3&   0.08  & --0.17 &  --0.16  & --1.16 &  --0.40\\  
CS\,29514-007 &  2.61 & 0.07~14  &   0.51&  2.73  & 0.08~5 & --0.11& 2.01  &  --0.56 & 2.59 & 0.18~3&  0.48  & 3.39&  0.07~2& --0.04  & --0.12 &  --0.23  & --0.96 &  --0.33\\  
CS\,29516-028 &  1.82 & 0.19~~~7 &   0.27&  2.07  & 0.08~3 & --0.22& 1.60  &  --0.42 & 2.17 & 0.01~2&  0.60  & 3.08&  0.02~2&   0.20  & --0.74 &  --0.31  & --1.98 &  --0.80\\  
CS\,30302-145 &  2.35 & 0.09~~~8 &   0.41&  2.67  & 0.10~5 & --0.01& 2.33  &  --0.08 & 2.57 & 0.04~2&  0.61  & 3.55&  0.09~2&   0.28  & --1.13 &  --1.09  &   --   &    --  \\  
CS\,30344-070 &  2.42 & 0.07~12  &   0.43&  2.61  & 0.02~5 & --0.12& 1.89  &  --0.57 &   -- &   -- &   --    & 3.23&  0.10~2& --0.09  & --0.17 &  --0.17  & --1.09 &  --0.35\\  
HE\,0148-2611 &  1.97 & 0.06~~~8 &   0.27&  2.27  & 0.05~3 & --0.17& 1.30  &  --0.87 &   -- &   -- &   --    & 2.83&  0.09~2& --0.20  & --1.54 &  --1.26  &   --   &    --  \\  
HE\,1413-1954 &  2.21 & 0.16~11  &   0.59&  2.60  & 0.35~2 &   0.25&   --  &    --   &   -- &   -- &   --    & 3.23&  0.21~2&   0.28  & --0.95 &  --0.58  & --1.10 &    0.01\\  
\hline                                          
\multicolumn{19}{l}{The standard deviation $\sigma$ is given when the abundance of the element is determined from more than two lines.}\\
\multicolumn{19}{l}{The abundances of Na, Al, and Ca have been corrected for NLTE.}\\
\multicolumn{19}{l}{The solar values of C and Fe are from \citet{abbosun}, while for the other elements, they are from \citet{lodders09}.}\\
\end{tabular}
\end{table*}

In Paper I, the stellar temperature was determined by different methods: photometry, and the profile of the wings of the hydrogen lines by using different theories for the self-broadening of H lines. The surface gravity was derived by enforcing the same Fe abundance from the \ion{Fe}{i} and \ion{Fe}{ii} lines. As usual, the microturbulence (\vt) was determined by ensuring that the abundance of iron is independent of the equivalent width of the lines.

\begin{figure}
\centering
\resizebox{6.0cm}{!}
{\includegraphics{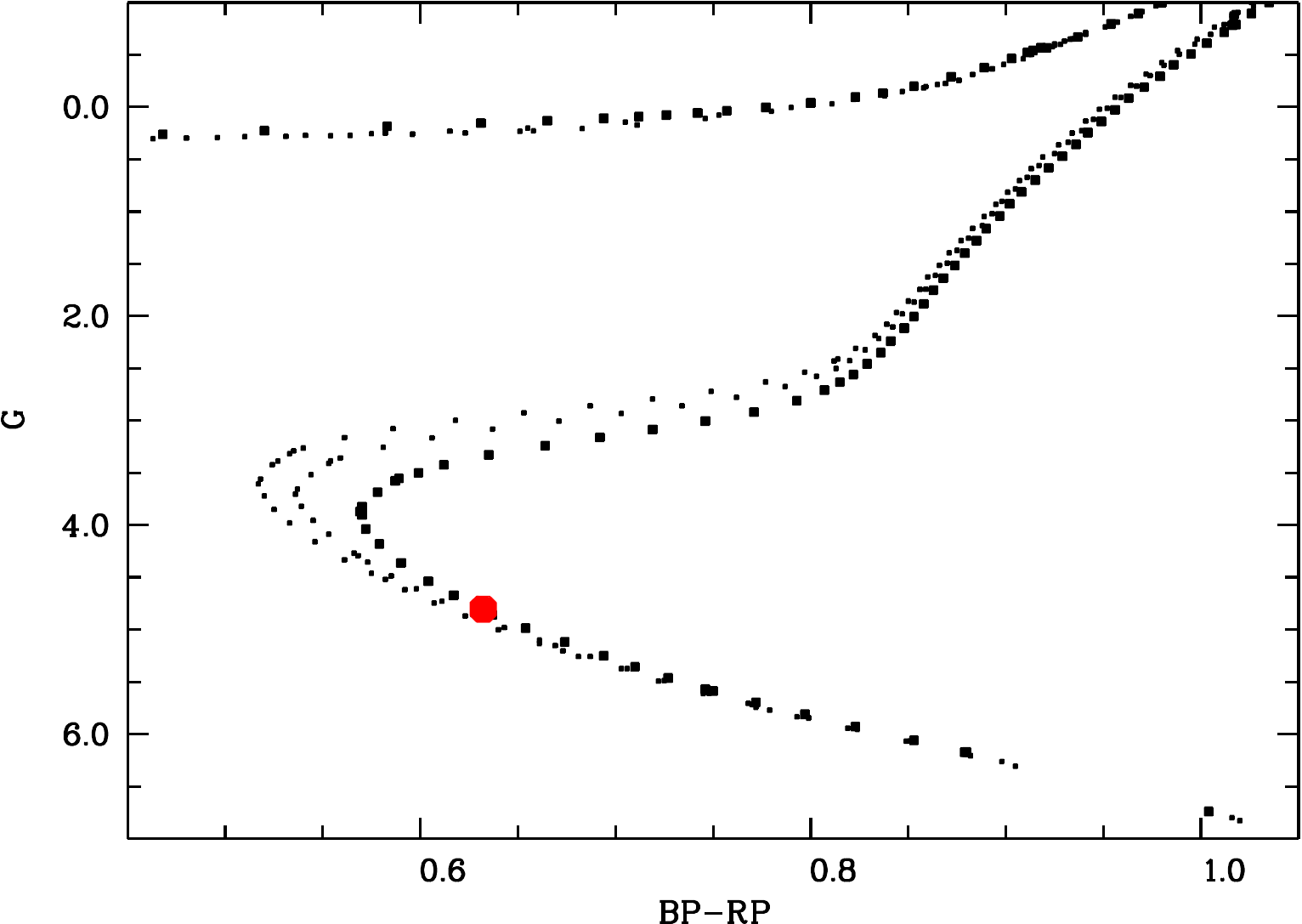}}
\caption{Comparison of the Gaia DR2 photometry of CS\,22188-033 to a PARSEC isochrone with a metallicity of --3.0 and age 8, 10, 12, and 14\,Gyr.}
\label{fig:isopar}
\end{figure}

In this investigation, we decided to take advantage of the information provided by the Gaia Data Release 2 \citep[Gaia DR2][]{Gaia,GaiaDR2,Arenou} to derive the stellar parameters: effective temperature and surface gravity.  
In this sample of turnoff stars, the precision of the parallaxes is very high, always better than 12\% and generally better than 5\%, which allows a very good determination of these parameters.
For consistency, we used the reddening E(B-V) given in Table\,2 of Paper\,I, deduced from the maps of \citet{schlegel98} and corrected as described in \citet{bonifacio00}. This reddening correction was converted into the Gaia photometric system by adopting the values derived for a G2V star recommended on the PARSEC site \footnote{\href{http://stev.oapd.inaf.it/cgi-bin/cmd}{http://stev.oapd.inaf.it/cgi-bin/cmd}.
Using the \citet{cardelli} plus \citet{odonnel} extinction curves with total-to-selective extinction ratio of 3.1 the following extinction coefficients are derived: 
$A_G/A_V = 0.85926$, $A_{BP}/AV = 1.06794$, $A_{RP} = 0.65199$}.   
The Gaia DR2 photometry is provided in Table\,\ref{tab:ParametrosEstelares}. 
By using the parallax provided by the Gaia\,DR2 catalogue and the reddening, we derived the absolute dereddenned G$_0$ 
magnitude and the dereddenned colour (BP--RP)$_{0}$. These values were then compared to 
the PARSEC isochrones \citep{Leo,marigo17} computed for different ages. 
Assuming the metallicity reported by \citet{sbordone10} for each star, we extracted from the PARSEC data base a grid
of isochrones with ages of 9 to 14 Gyr, with Gaia DR2 colours, computed using the
passbands of \citet{evans}. We then interpolated for the observed (BP--RP)$_{0}$ colour and
the absolute $G$ magnitude 
for each age, and selected the age for which this value was closest
to the observed absolute $G$ magnitude.  At this point, we obtained
\teff\ and \logg\ by interpolating in the isochrone. 
The  \teff\ and \logg\ values of the stars deduced from this comparison are given in 
Table\,\ref{tab:ParametrosEstelares}. In Fig.\,\ref{fig:isopar}, an example of the fit 
of the isochrones is given for the star CS\,22188--033. 
We estimate that the uncertainty in \teff\ does not exceed 100\,K and is about 0.3\,dex in \logg.
 
\begin{figure}
\centering
\resizebox{5.4cm}{!}
{\includegraphics{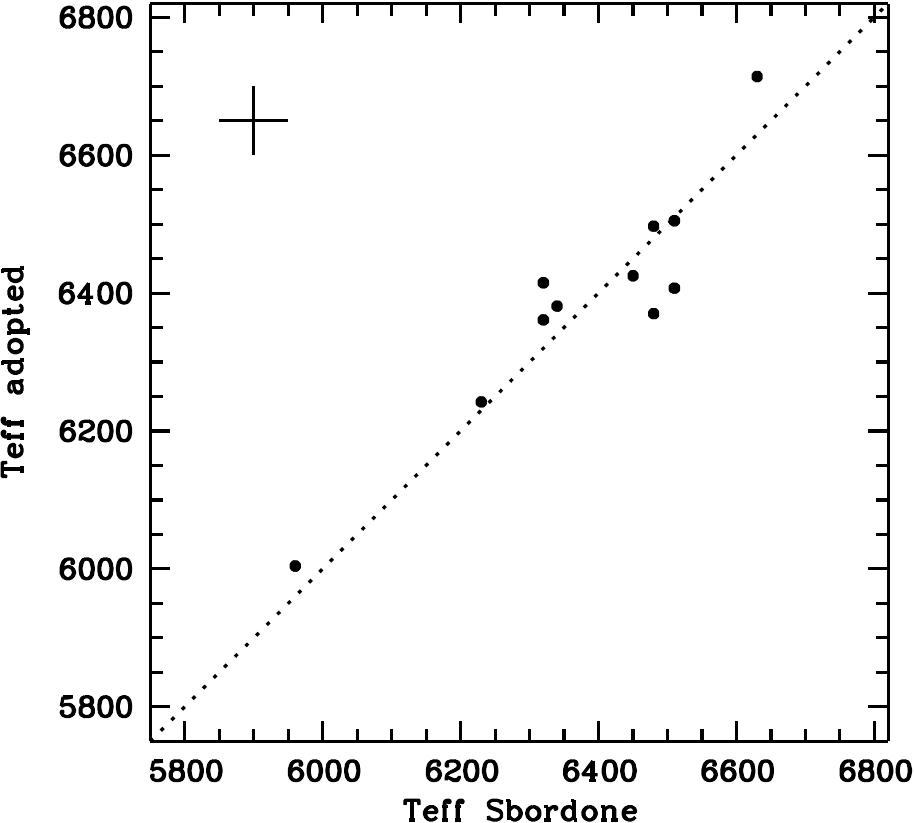}}
\resizebox{5.0cm}{!}
{\includegraphics{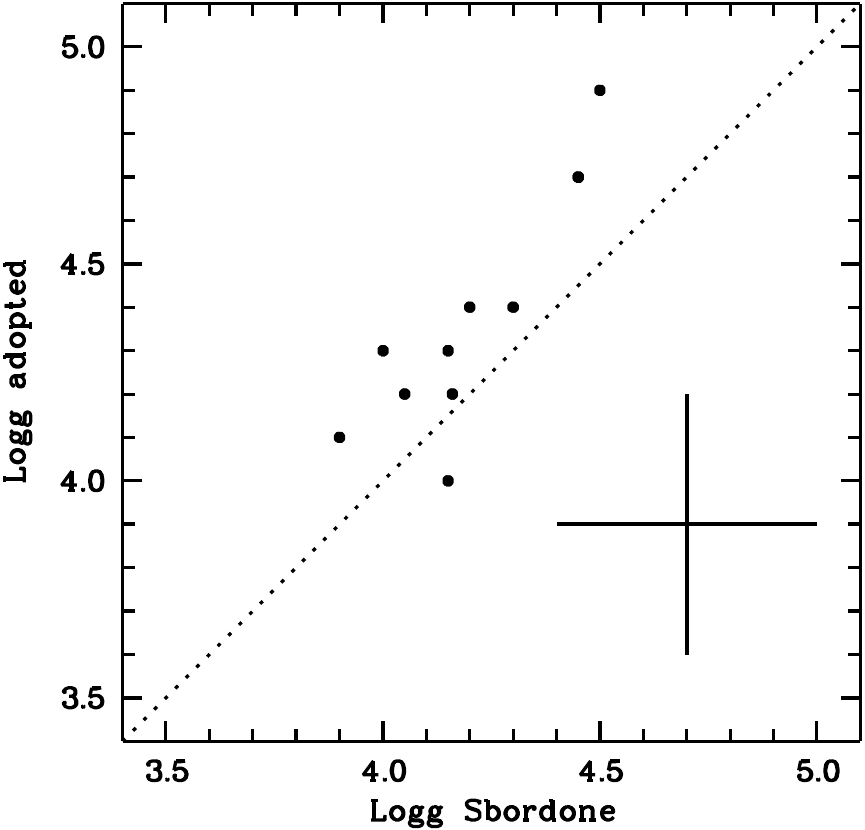}}
\caption{Comparison of the \teff\ and \logg\ we adopted and the values of the 3D model adopted in Paper I.}
\label{fig:compmd1}
\end{figure}

In Fig.\,\ref{fig:compmd1} we compare our adopted parameters \teff\ and \logg\ (Table\,\ref{tab:ParametrosEstelares}) to those adopted in Paper\,I. The agreement is good within the uncertainties ($\rm \Delta(T_{eff})= 8\pm 67$\,K,
$\rm \Delta(log g) = -0.17\pm 0.14$\,dex, where the differences are ours; Paper\,I). Our \logg\ value is systematically higher by about 0.1\,dex than the value adopted in Paper\,I, however. This may reflect the fact that departures from  local thermodynamic equilibrium (LTE) were not taken into account in the calculation  of the ionisation equilibrium of Fe in Paper\,I.
Following \citet{TheveninIdiart99}, this error may sometimes reach 0.5\,dex, in particular, if the low-excitation potential lines of \Feu, which are very sensitive to non-LTE effects, are used for this determination (but this was  not the case in Paper I).\\ 
When our microturbulence velocities are compared to the values adopted in Paper I, the differences never exceed 0.2\,\kms. This value corresponds to the uncertainty of the determination.

\subsection{Abundance determination}
A  detailed chemical analysis based on the hypothesis of LTE was performed.  MARCS model atmospheres \citep{GustafssonBE75,GustafssonEE03,Plez08} were used with the code Turbospectrum \citep{alvarez98,Plez-code12}, which computes synthetic spectra to be compared to the observations, allowing us to carry out the study of the element abundances.\\
 When the lines are severely blended or double (e.g. molecular lines or the Li feature), when the damping wings are important (strong Mg lines), or when the lines are affected by hyperfine structure like the \Bad~ lines, the determination of the abundance was conducted by spectral synthesis. Otherwise, we relied on equivalent width measurements.\\
The abundances of the elements are given in Table\,\ref{tab:Abundance}. (As usual, for each element X, A(X)= log(N(X)/N(H))+12,  $\rm[X/H]=A(X)-A(X)_{\odot}$, and [X/Fe]=[X/H]--[Fe/H].) The iron abundances deduced from the \Feu~and \Fed~ lines are given in Table \ref{tab:ParametrosEstelares}. The value of [Fe/H] in Table\,\ref{tab:Abundance} is the mean of these values.

To calculate [X/Fe], we used the solar abundance values $\rm A(X)_{\odot}$ provided by \citet{abbosun} when available, otherwise we adopted those by \citet{lodders09} (see Table\,\ref{tab:Abundance}).

Because our sample of stars has been observed and analysed in almost the same conditions as the stars of the ESO Large Programme ``First Stars, first nucleosynthesis'' (First Stars, for short), we directly compare our results in the subsequent sections to those of \citet{bonifacio09} and \citet{cayrel04}. However, we must note that the quality of the spectra obtained for these 11 stars is not as good as the quality of the spectra used in \citet{cayrel04} and \citet{bonifacio07,bonifacio09}. The S/N is generally lower than half the S/N obtained for these preceding papers. This parameter is very important in case of very or extremely metal-poor stars, where the lines of many elements are often very weak. In particular, we do not intend to study the star-to-star scatter of the different elements as a function of the metallicity, as was done in the previously cited papers, but only to try to detect some systematic abundance anomalies in the Li-poor stars.

\subsection{Abundance uncertainties}
Most of the lines of the elements we studied, between C to Ba, are located in the wavelength range 370 -- 550\,nm. In this region,the S/N of the spectra is only about half of the S/N obtained in the region of the Li feature (671\,nm).  As a consequence, the first cause of error for these elements is the uncertainty in the equivalent width (or the profile) of the lines. This uncertainty is generally dominated by the uncertainty in the position of the continuum. In case of weak lines (EW < 10 mA), when the S/N of the spectra is poor, the uncertainty on the equivalent width can be as large as a factor of two, which results in an error 0.3\,dex on A(X). This is particularly the case for the lines of HE1413-1954. A second cause of error is the uncertainties in the \gf~ values that are adopted.\\
The resulting error  is estimated as $\rm\sigma / \sqrt(N-1)$ (where N is for each element X the number of the lines and $\sigma$ the rms around the mean abundance). In a conservative way, when there are only two lines, an error of 0.15\,dex was adopted even if it was $\sigma < 0.15$\,dex. In case of only one line, we adopted an error of 0.2 dex. 

To determine the uncertainties in the abundance determinations A(X) or [X/H], the uncertainties in \teff, \logg,\ and microturbulence \vt~ must be also considered.
For these turnoff metal-poor stars, the adopted uncertainties on \teff, \logg,~ and \vt~ are 100\,K, 0.3\,dex, and 0.2\,\kms (see Sec\,\ref{sec:stelpar}). Following \citet{bonifacio09}, these values correspond to an uncertainty on [Fe/H] of about 0.11\,dex. 
We also computed the error on A(Li) due to the uncertainty in the atmospheric parameters (not given in \citealt{bonifacio09}). For turnoff stars, this error is about 0.07\,dex, and it is dominated by the error on the effective temperature. 
The total error on A(X) or [X/H] includes errors linked to the choice of stellar parameters of the model and to observations.

For elements X heavier than Li, the interesting value is the ratio [X/Fe]. For most of the elements, an error on, for example, the temperature of the model induces about the same error on A(Fe) and A(X), and as a consequence, the error on [X/Fe] is small. We have adopted the values given in the Table 4 of \citet{bonifacio09}, taking into account (linearly) that in our case the adopted error on log g is three times larger.

\begin{table}
\centering
\scalefont{0.8}
\caption{Lithium abundances.}
\label{tab:Li}
\begin{tabular}{lccccccc}
\hline
\hline\\
Star         &      & EW(Li) &  $A$(Li)  &           & $A$(Li)    \\
             & S/N  & [pm]   & 1D LTE    &   error   & 3D NLTE    \\
\hline\\
BS\,17572-100& 190  &\hspace*{2.2mm}$1.71$ &\hspace*{2.2mm}$2.19$   &  $0.08$  &\hspace*{2.2mm}$2.16$ \\
CS\,22188-033*& 220 &\hspace*{2.2mm}$0.90$ &\hspace*{2.2mm}$1.74$   &  $0.08$  &\hspace*{2.2mm}$1.72$ \\
CS\,22882-027&  80  & $<0.56$              & $<1.60$                &  $0.10$  & $<1.56$              \\
CS\,22950-173&  90  &\hspace*{2.2mm}$2.25$ &\hspace*{2.2mm}$2.23$   &  $0.10$  &\hspace*{2.2mm}$2.20$ \\
CS\,29491-084& 100  &\hspace*{2.2mm}$1.86$ &\hspace*{2.2mm}$2.15$   &  $0.10$  &\hspace*{2.2mm}$2.13$ \\
CS\,29514-007&  90  &\hspace*{2.2mm}$2.49$ &\hspace*{2.2mm}$2.27$   &  $0.10$  &\hspace*{2.2mm}$2.25$ \\
CS\,29516-028&  60  &\hspace*{2.2mm}$2.46$ &\hspace*{2.2mm}$2.01$   &  $0.13$  &\hspace*{2.2mm}$2.02$ \\
CS\,30302-145&  70  &\hspace*{2.2mm}$1.59$ &\hspace*{2.2mm}$2.17$   &  $0.12$  &\hspace*{2.2mm}$2.14$ \\
CS\,30344-070&  80  &\hspace*{2.2mm}$1.70$ &\hspace*{2.2mm}$2.22$   &  $0.12$  &\hspace*{2.2mm}$2.19$ \\
HE\,0148-2611*& 200 &\hspace*{2.2mm}$1.26$ &\hspace*{2.2mm}$2.08$   &  $0.08$  &\hspace*{2.2mm}$2.05$ \\
HE\,1413-1954&  50  &\hspace*{2.2mm}$1.62$ &\hspace*{2.2mm}$2.17$   &  $0.13$  &\hspace*{2.2mm}$2.14$ \\
\hline                                                
\multicolumn{6}{l}{ An asterisk means that new spectra have been obtained.}\\
\end{tabular}
\end{table}

\begin{figure}
\centering
\resizebox{8.0cm}{!}
{\includegraphics{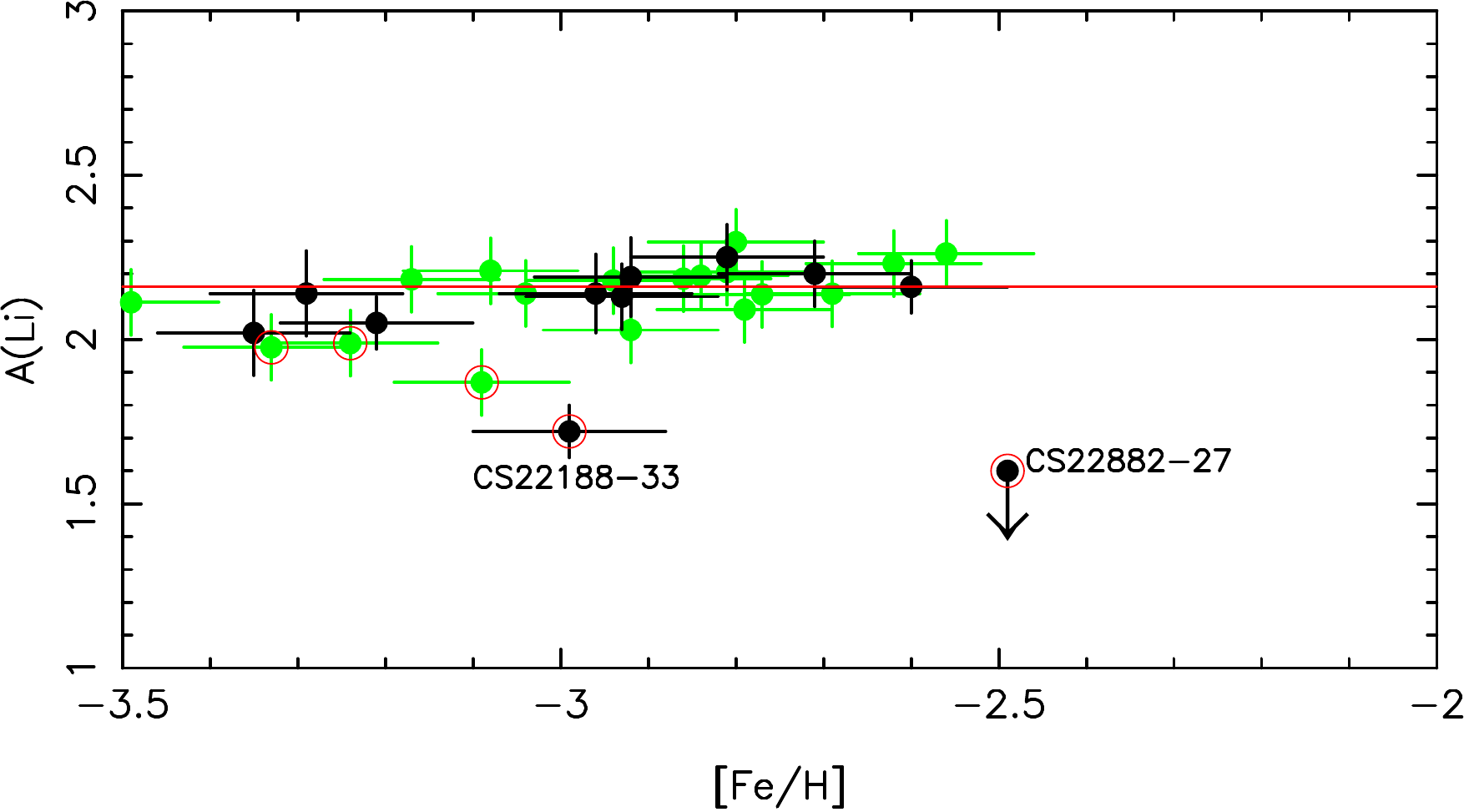}}
\caption{Lithium abundance vs. [Fe/H] for the stars of our sample (black dots)  and of the  First Stars sample (green dots). The plateau is indicated by a red line. In both samples, the stars that deviate by more than 2$\sigma$ from the plateau are surrounded by a red circle.}
\label{fig:abli}
\end{figure}

\begin{figure*}
\centering
\resizebox{6.5cm}{!}
{\includegraphics{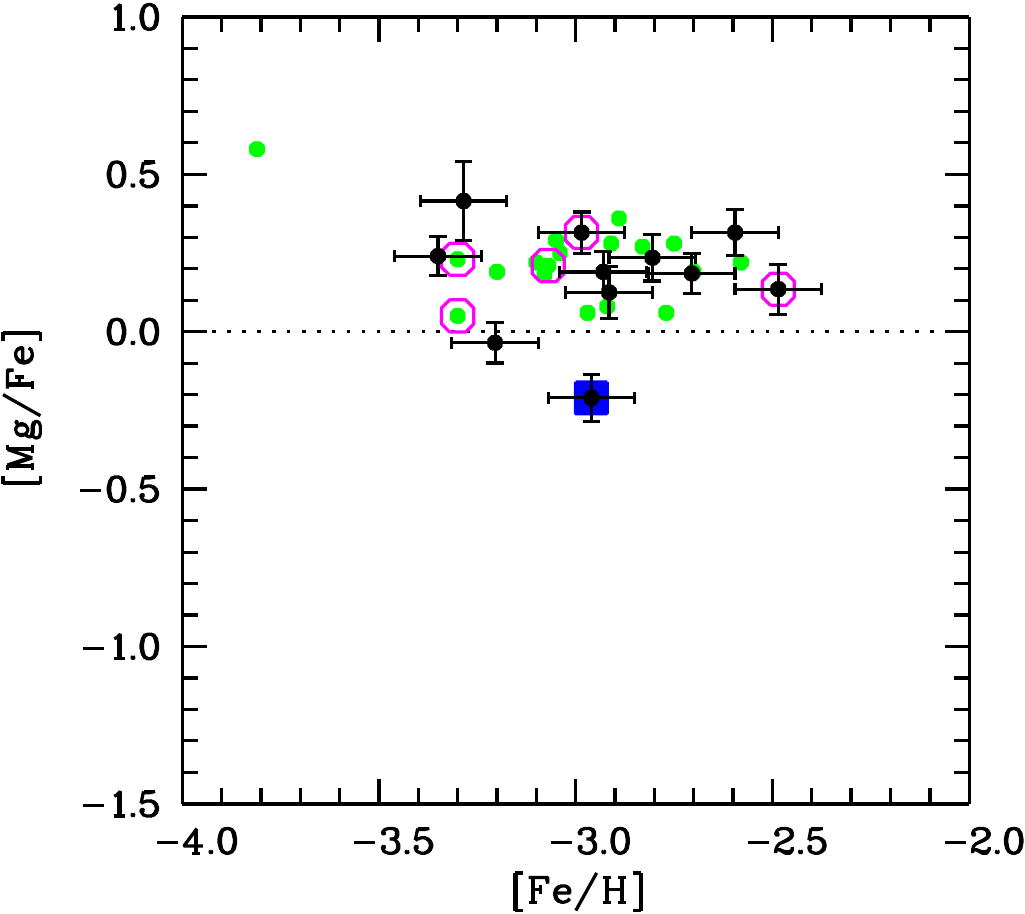}}
\resizebox{6.5cm}{!}
{\includegraphics{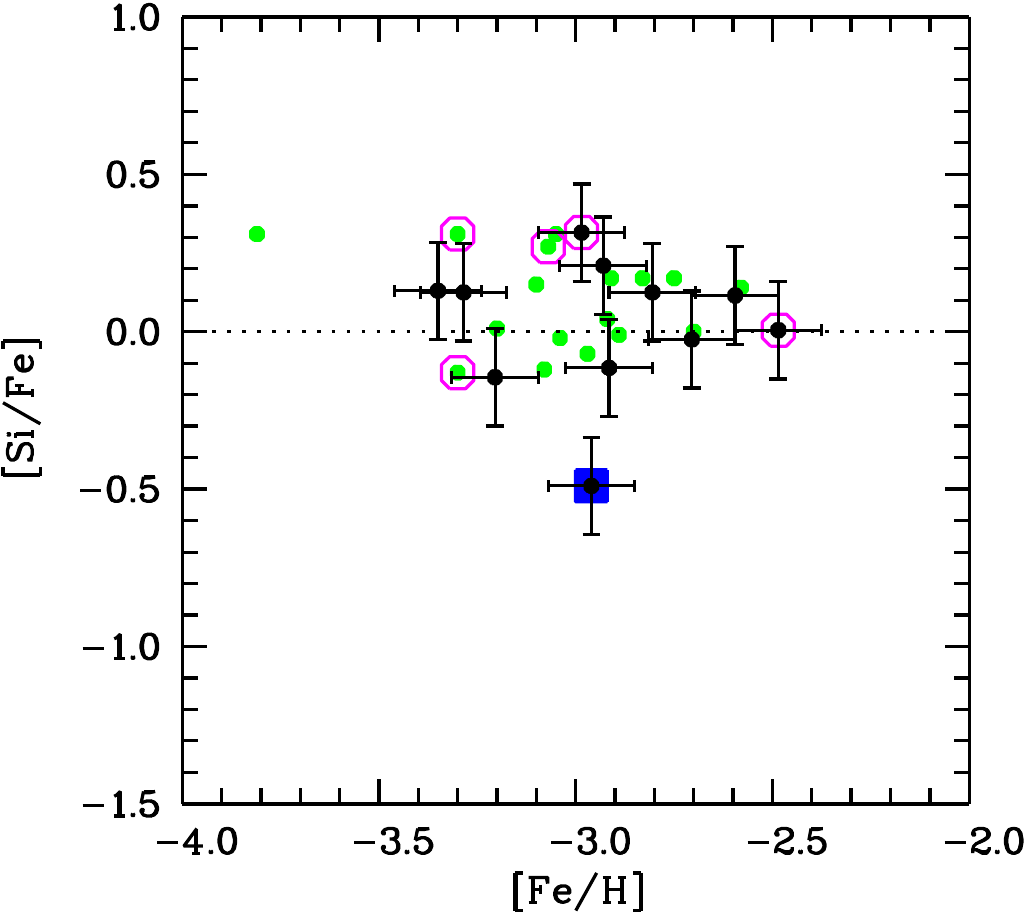}}
\resizebox{6.5cm}{!}
{\includegraphics{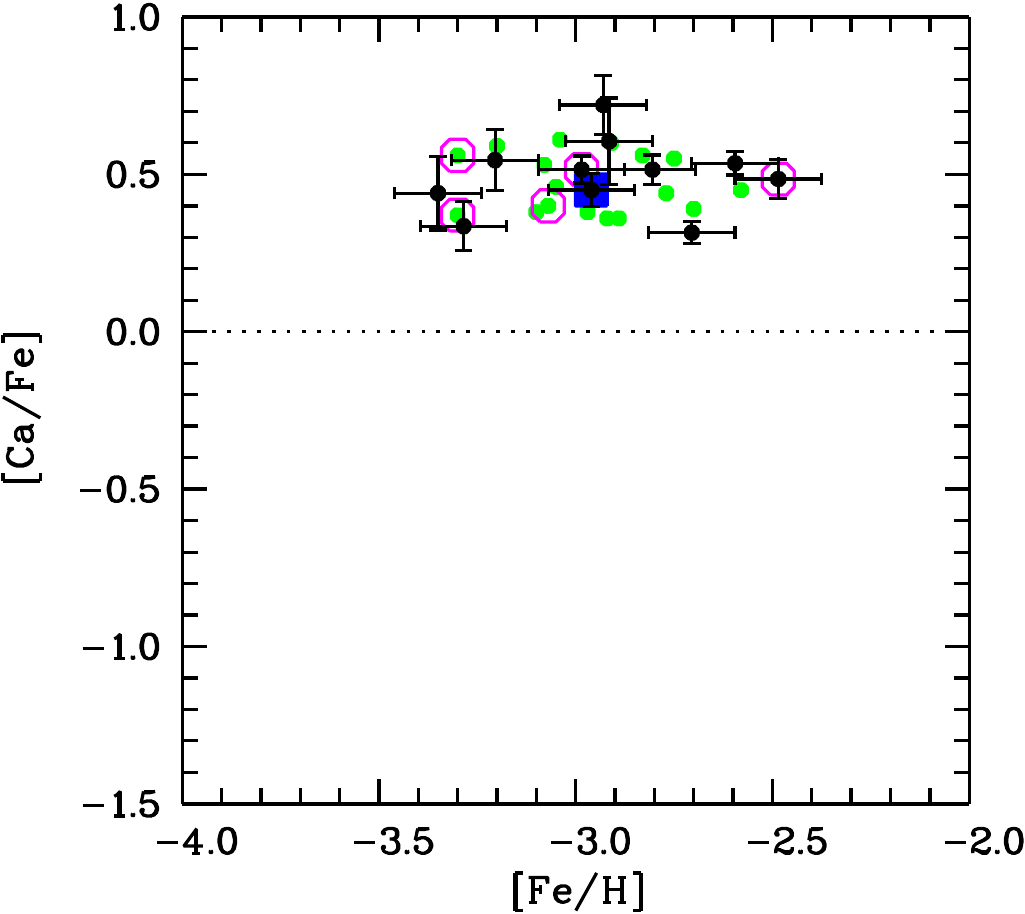}}
\resizebox{6.5cm}{!}
{\includegraphics{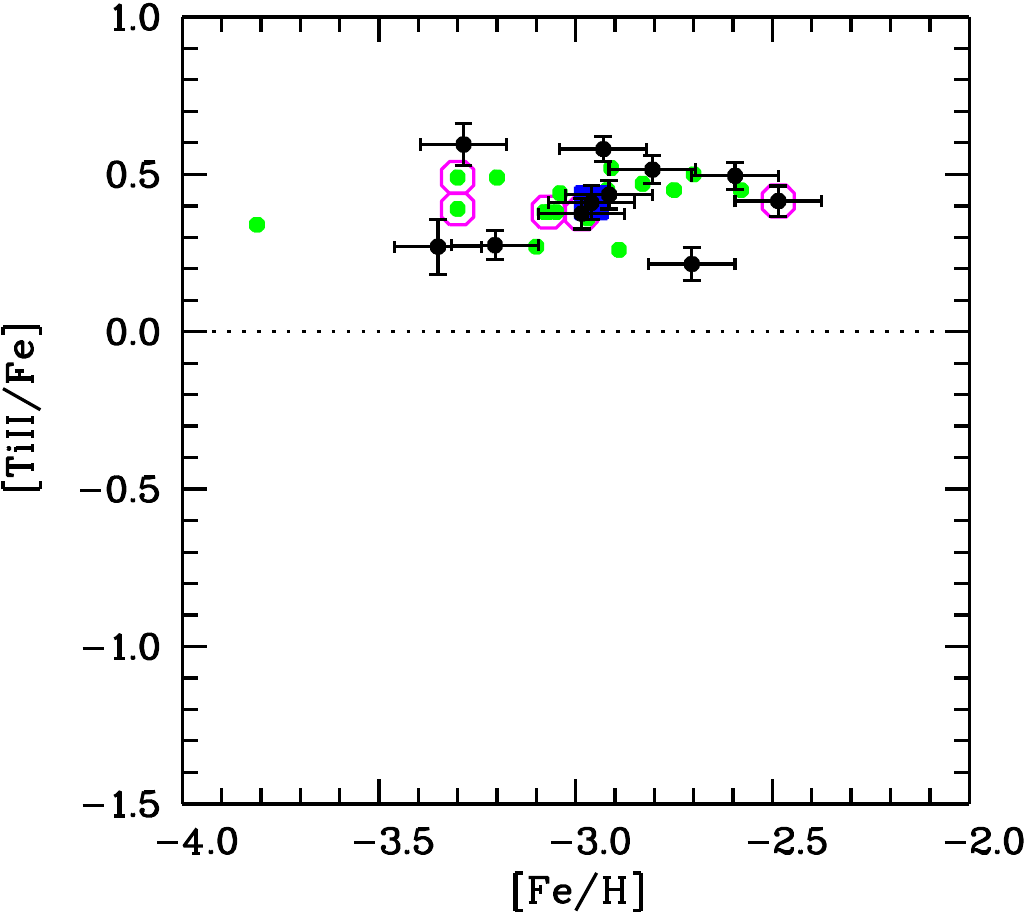}}
\caption{Abundance ratios of the $\alpha$-elements as a functions of [Fe/H]. The black dots refer to the present study,  and the green dots to the First Stars sample. The  very lithium-poor stars are surrounded by a pink circle. The $\alpha$-poor star CS\,30302-145 is represented by a blue square.}
\label{fig:alpha}
\end{figure*}

\begin{figure*}
\centering
\resizebox{6.5cm}{!}
{\includegraphics{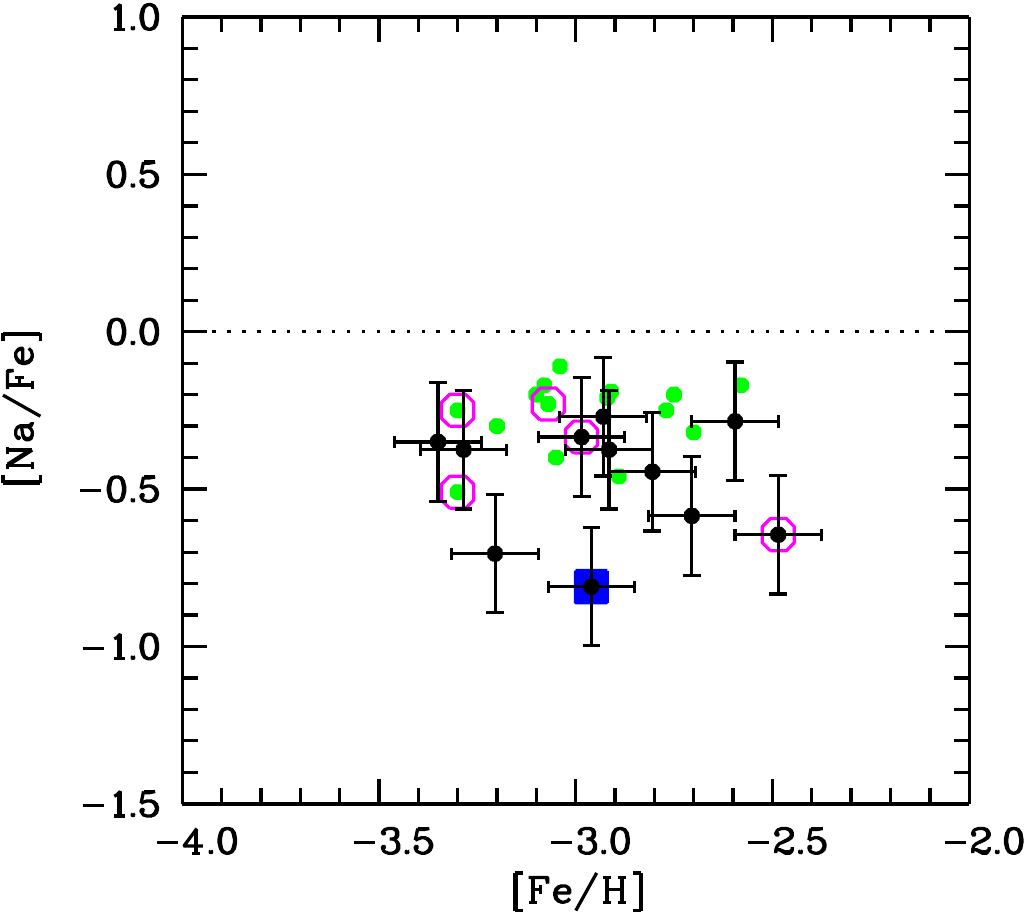}}
\resizebox{6.5cm}{!}
{\includegraphics{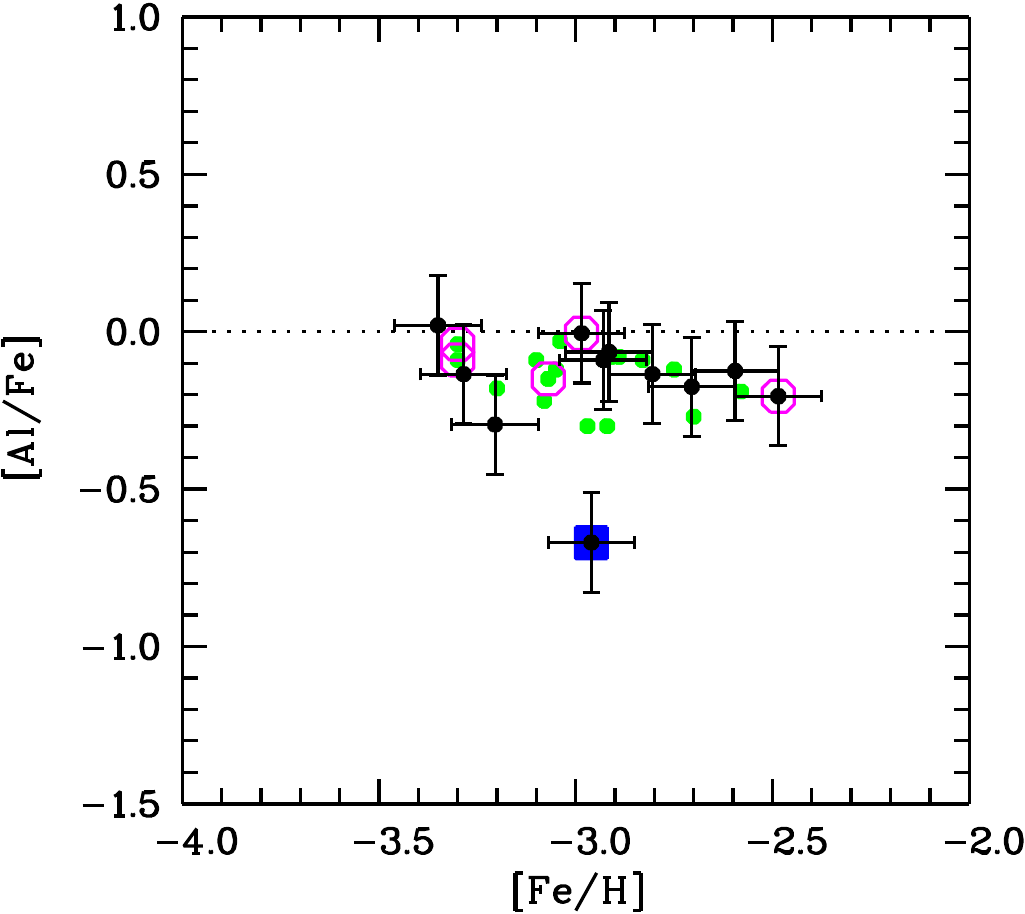}}
\caption{Odd element abundance. Symbols are the same as in figure.\,\ref{fig:alpha}.}
\label{fig:odd}
\end{figure*}

\begin{figure*}
\centering
\resizebox{6.5cm}{!}
{\includegraphics{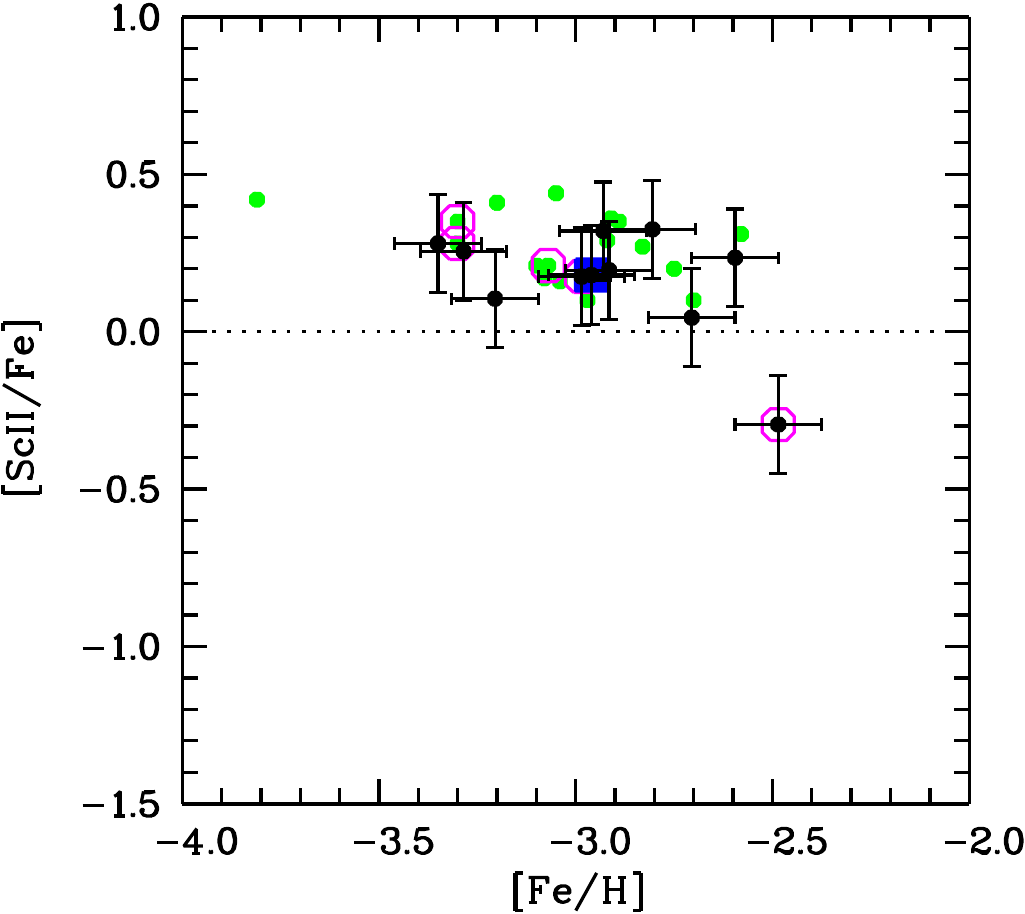}}
\resizebox{6.5cm}{!}
{\includegraphics{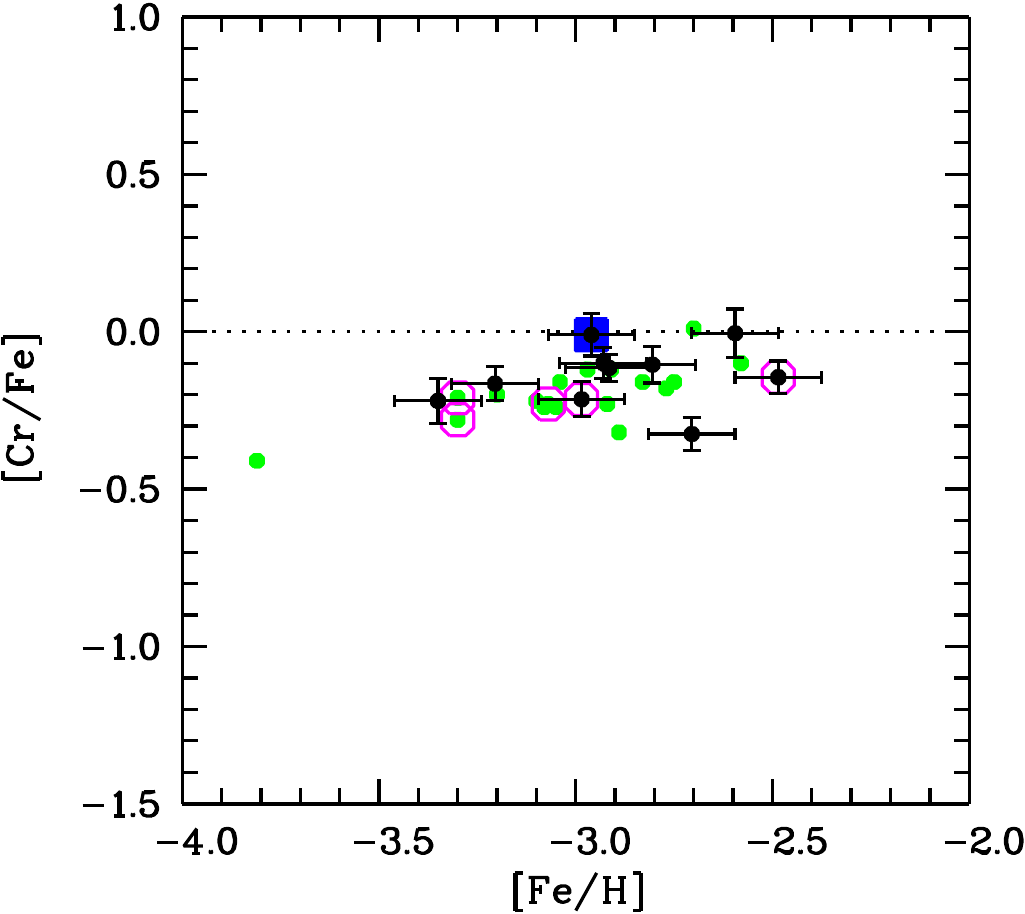}}
\resizebox{6.5cm}{!}
{\includegraphics{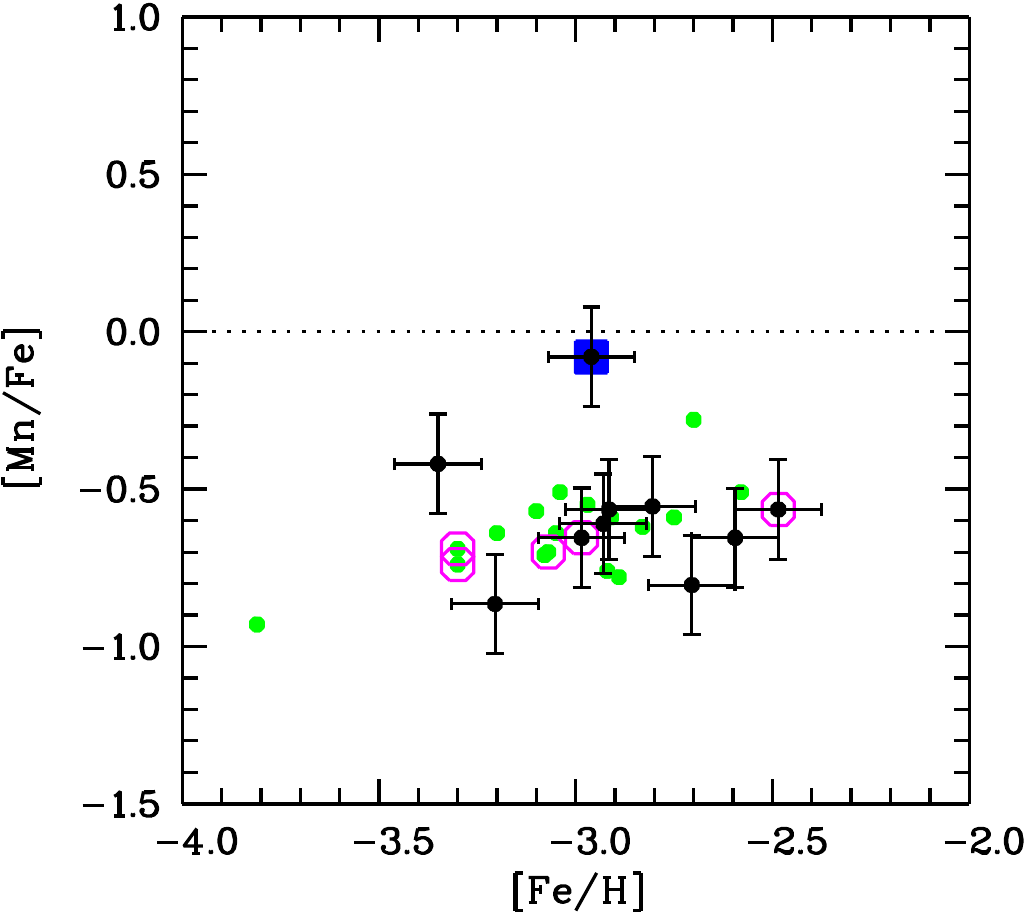}}
\resizebox{6.5cm}{!}
{\includegraphics{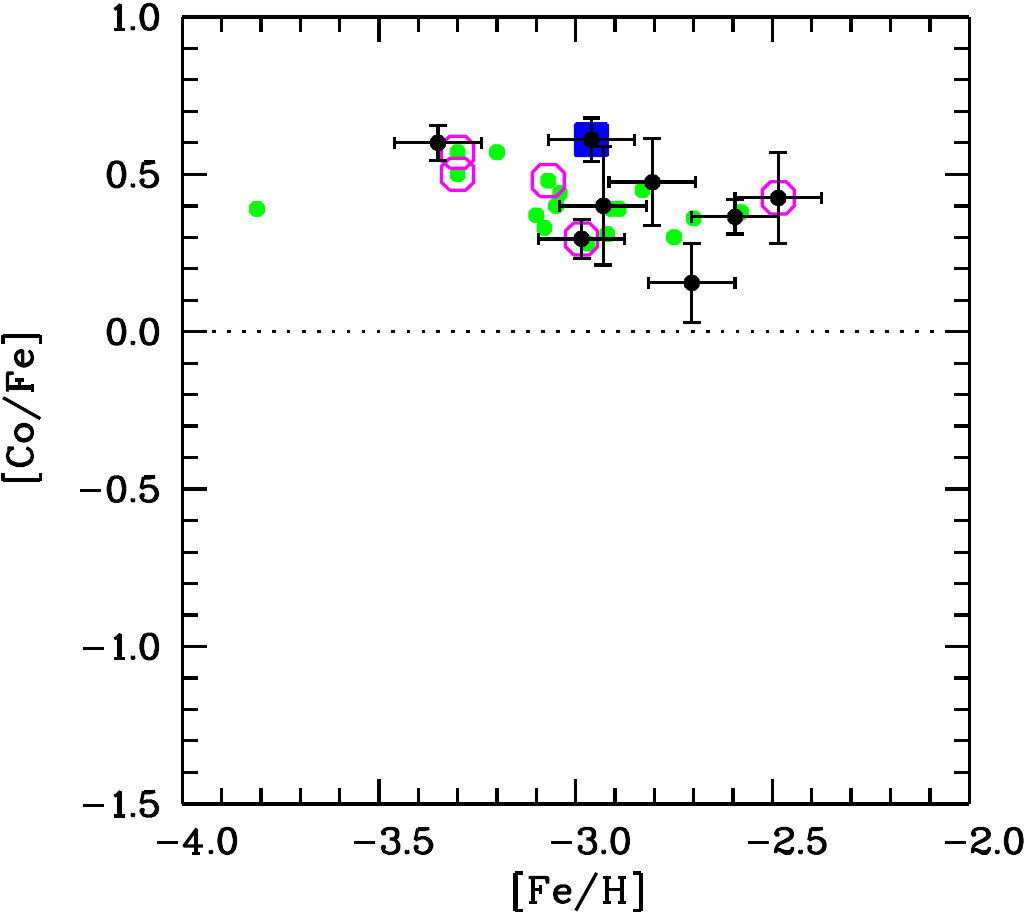}}
\resizebox{6.5cm}{!}
{\includegraphics{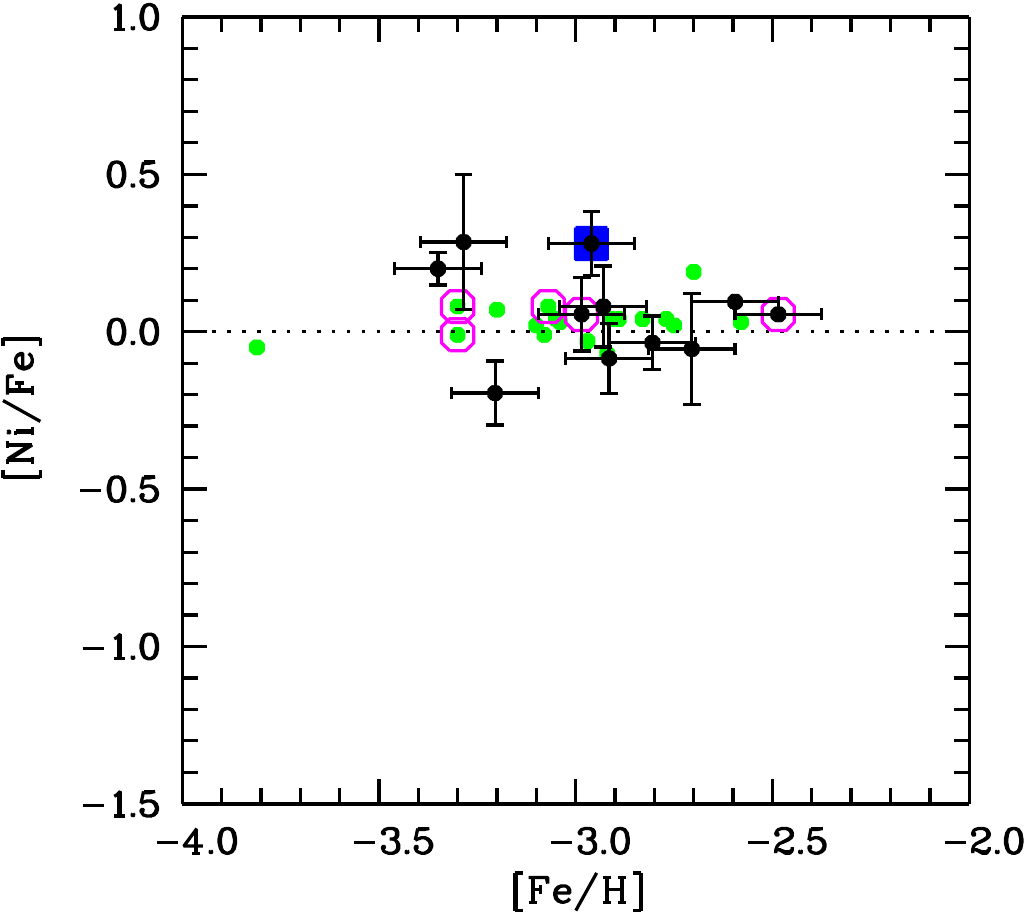}}
\caption{Iron peak elements. Symbols are the same as in figure.\,\ref{fig:alpha}.}
\label{fig:ironpeak}
\end{figure*}

\begin{figure*}
\centering
\resizebox{6.0cm}{!}
{\includegraphics{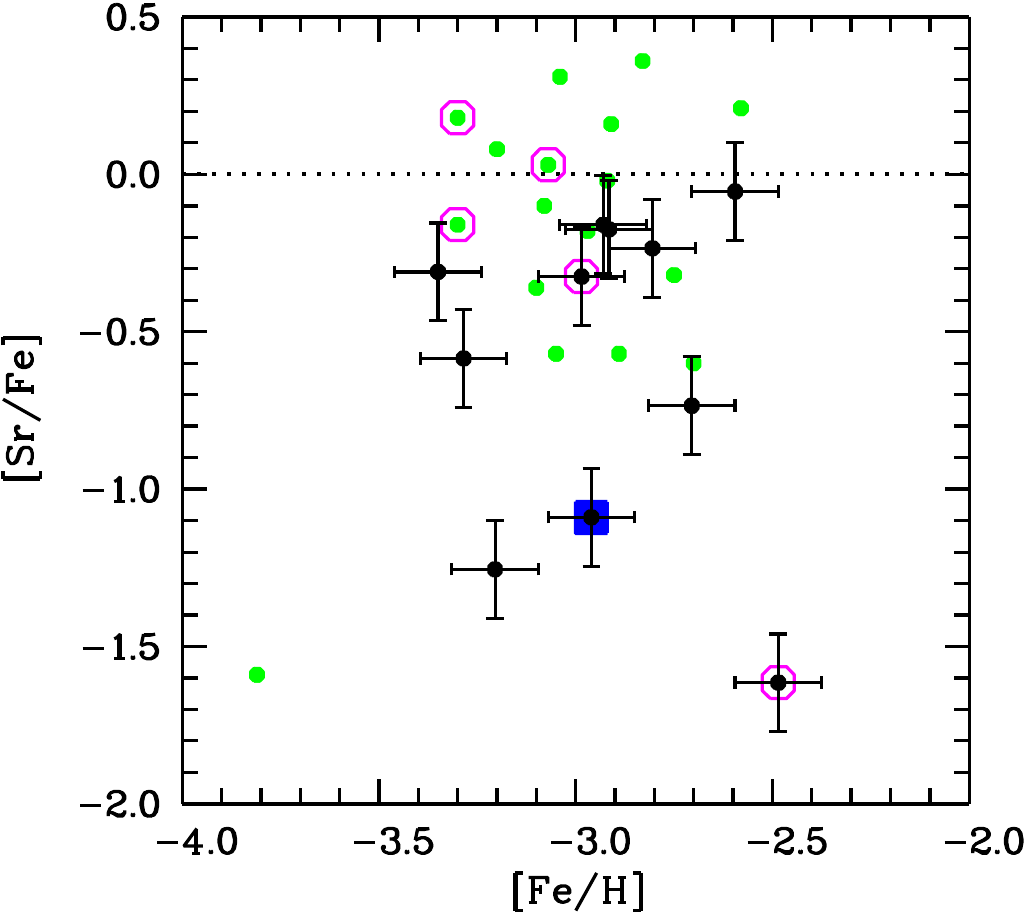}}
\resizebox{6.0cm}{!}
{\includegraphics{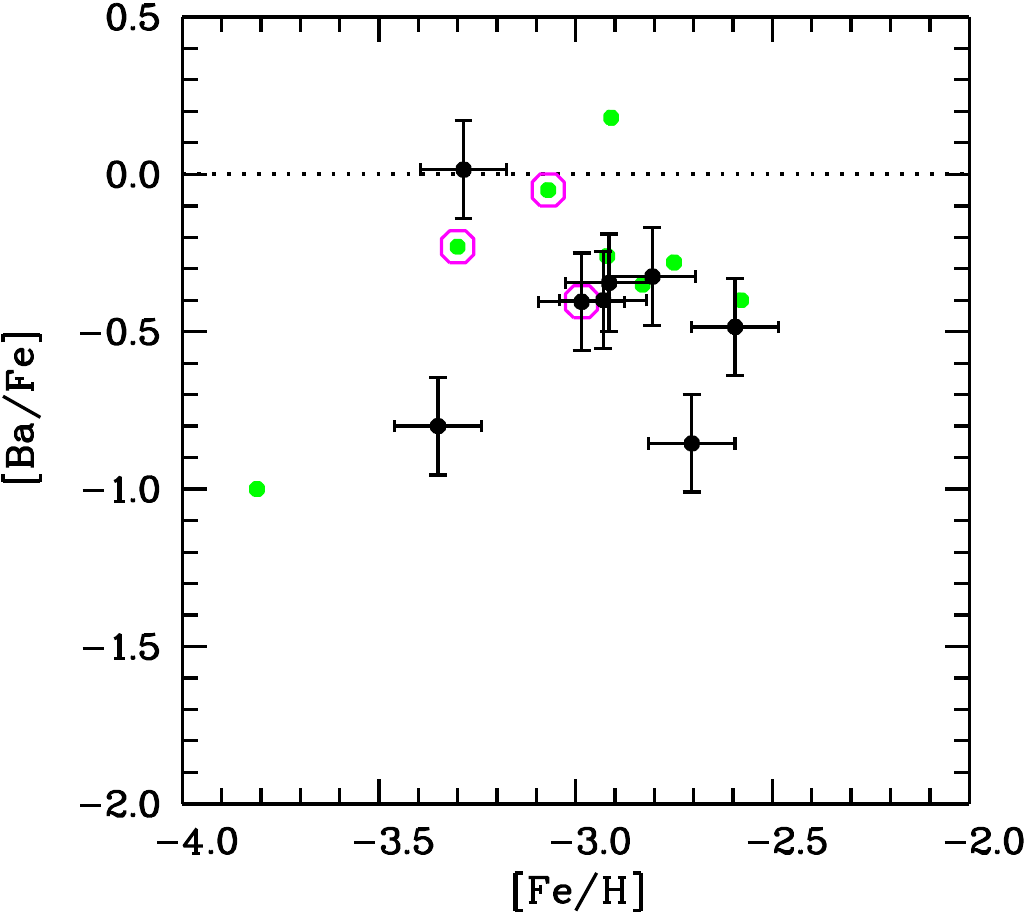}}
\caption{Neutron-capture elements. Symbols are the same  as in figure.\,\ref{fig:alpha}.}
\label{fig:heavy}
\end{figure*}

\subsection{Lithium}   \label{sec:liab}
\subsubsection{1D LTE computation}
The lithium analysis was first conducted by spectral synthesis, investigating the Li 670.7\,nm resonance doublet.
The adopted atomic data for the Li doublet are the same as those used by \citet{bonifacio07} and \citet{sbordone10}; they are based on  \citet{AsplundLN06}. They take into account the hyperfine structure of the lines and also the isotopic components for an assumed solar isotopic ratio. The abundance A(Li), based on one-dimension models with the approximation of local thermodynamical equilibrium (1D-LTE), is given in Col. 3 of Table\,\ref{tab:Li}.

\subsubsection{3D NLTE correction}

Abundance corrections due to 3D and NLTE effects may be taken into account
by using the analytic fitting functions developed by \cite{mott20}. Using
the individual stellar parameters from Table\,\ref{tab:ParametrosEstelares},
the metallicity [Fe/H] from Table\,\ref{tab:Abundance}, 
and the 1D-LTE lithium abundance $A$(Li) from Table\,\ref{tab:Li} as input
for their function {\tt FFI}, we obtain corrections in the range
$-0.01$\,$\la$\,$\Delta($FFI$)$\,$\la$\,$+0.04$ (assuming
$^6$Li/$^7$Li\,=\,$0$). This closely agrees with the corrections for
stars on the Spite plateau shown in Fig.\,16 of \cite{mott20}.

Alternatively, we can use fitting function {\tt FFIII} with the equivalent
width EW from Table\,\ref{tab:Li} instead of the 1D LTE Li abundance. With
this method, we obtain corrections in the range
$-0.05$\,$\la$\,$\Delta($FFIII$)$\,$\la$\,$-0.03$, significantly more negative
than the corrections derived from {\tt FFI}.

This disagreement arises because the treatment of the radiative
opacities in the 3D CO5BOLD \citep{freytag12} model atmosphere and the 1D MARCS model differs in its level of detail (CO5BOLD opacity groups versus
detailed MARCS opacity sampling).  The corrections $\Delta($FFI$)$ are valid
if the difference between $A$(Li)$_{\rm 3D-NLTE}$\,$-$\,$A$(Li)$_{\rm 1D-LTE}$
is insensitive to the description of the opacities as long as the same
treatment is used to construct the 3D and the 1D model. On the other hand,
$\Delta($FFIII$)$ is valid if the EW of the 3D NLTE synthetic line profile
is insensitive to small changes in 3D temperature structure related to
the description of the radiative opacities.

Comparison of 1D models with different opacity schemes shows that neither of
the above conditions is strictly valid \citep[see e.g.\,Fig.\,15 in][]{mott20}.
A better approximation is to assume that
$A$(Li)$_{\rm 3D-NLTE}$\,$-$\,$A$(Li)$_{\rm 1D-NLTE}$ is invariant under an
identical change of the opacities in the 3D and the 1D model. In this case,
the abundance correction is given by $\Delta_3$\,=\,$\Delta($FFI$) +
\Delta_{\rm 1D}^{(0)} - \Delta_{\rm 1D}^{(1)}$\,, where $\Delta_{\rm 1D}^{(0)}$ and
$\Delta_{\rm  1D}^{(1)}$ denote the NLTE corrections computed from 1D models
with CO5BOLD and OSMARCS opacities, respectively, but otherwise identical
stellar parameters. We find that $\Delta_{\rm 1D}^{(0)} - \Delta_{\rm 1D}^{(1)}$
is approximately $-0.03$, and hence $\Delta_3$
ranges between $-0.04$ and $+0.00$ for our sample of stars.

We note that the Li abundance corrections for stars on the Spite plateau
provided by \cite{wang20} appear to be slightly more negative than our
$\Delta($FFI$)$ by about $0.003$\,dex, judging from their Fig.\,11,
presumably owing to a different treatment of the UV background opacities
in calculating the lithium line formation in statistical equilibrium.
Nevertheless, their raw corrections agree  almost perfectly with
our refined (final) corrections $\Delta_3$.\\

\subsubsection{Li abundance}
The 3D NLTE Li abundances given in the last column of Table\,\ref{tab:Li} are
computed as $A$(Li)$_{\rm 1D-LTE} + \Delta_3$.\\
In Fig. \ref{fig:abli} we plot the 3D NLTE Li abundance versus metallicity for our sample of stars (black dots) and for the stars of the  First Stars (green dots).
 
The observational error is a function of the S/N of the spectrum. We estimated that in this region of the spectrum, it is close to 0.06\,dex for S/N=100 and 0.10\,dex for S/N=70. These estimates are based on the Cayrel formula \citep{Cayrel1988} and on a direct evaluation of the uncertainty of the best fit of the observed profile because the Cayrel formula does not take the uncertainty in the position of the continuum into account in particular. 
 The estimation of the total error on the  Li abundance (quadratic sum of the uncertainty of the model (0.07 dex) and of the observational error is given in Col. 4 in Table \ref{tab:Li}. 
In Fig. \ref{fig:abli}, several stars deviate significantly from the average. We drew a red circle around the stars whose Li abundance deviates from the average by more than twice the total error given in Col. 4 in Table \ref{tab:Li}. When these stars are not taken into account, we find that for our sample, the level of the plateau   is at $\rm A(Li)=2.14 \pm 0.07$ and at $\rm A(Li)=2.16 \pm 0.07$ when we also consider the  First Stars sample. The level of this plateau is within the uncertainty compatible with the values found by \citet{bonifacio07} (A(Li)=2.10) and in Paper I~ (A(Li)$\rm_{3D-NLTE}$=2.20). In our sample, only two stars have a Li abundance that deviates from the plateau by more than  $2\sigma$: CS\,22188-033 and CS\,22882-027.

The star CS\,22188-033 ([Fe/H] = --2.99)  shows a significant lithium depletion (Fig.\ref{fig:abli}). 
It confirms a meltdown of the Spite plateau  below [Fe/H]=--2.9. \\ 
Within the  First Stars, three other stars with [Fe/H]<--2.9 were also found with a Li abundance deviating by more than 2 $\sigma$ from the plateau:  CS\,22888-031, CS\,22948-093, and CS\,22966-011 \citep[][ and Paper I]{bonifacio07}.

In CS\,22882-027, the lithium line is not visible, and it was only possible to derive an upper limit of A(Li). This star is the least metal-poor star of our sample ([Fe/H]= --2.5). In the following, we say for convenience that this star is lithium free. It is discussed  in Section \ref{sec:age}.

\subsection{Carbon and nitrogen}
In this study, C and N abundances were derived from the analysis of  molecular bands: for C, the G-band at 425-450\,nm formed by CH lines, and for N, the NH band around 336\,nm.
The parameters of the G-band \citep{MasseronPVE14}, and the NH band ({\tt \footnotesize http://kurucz.harvard.edu/linelists/linesmol}) were directly taken from the website of Bertrand Plez \footnote{\url{https://nextcloud.lupm.in2p3.fr/s/r8pXijD39YLzw5T}}.\\
For all stars in our sample, we analysed the G-band around 430\,nm. It is rather weak in these very  metal-poor and relatively hot stars, and generally, only an upper limit could be derived. This is provided in Table\,\ref{tab:Abundance}.

The NH band in the near-UV (around 336\,nm) is generally outside the spectral range covered by our data. This wavelength range was observed in only two stars: BS\,17572-100 and HE\,1413-1954. The NH band could be measured in HE\,1413-1954, but it is not visible in BS\,17572-100 (as is generally the case in turnoff stars).

In our sample, one star, HE\,1413-1954, has been reported to be a carbon-enhanced metal-poor (CEMP) star \citep{masseron10,pols12}. We confirm the high abundance of C in this hot turnoff star:  [C/Fe]=+1.7\,dex (Table \ref{tab:Abundance}). 
From the NH band, we found a very high N abundance (A(N)=7.3, i.e. [N/Fe]=2.7\,dex) in agreement with the literature (see  \citet{pols12,LucatelloBC06}).\\ 
In metal-poor stars, the proportion of CEMP stars increases when the metallicity decreases. Following \citet{LucatelloBC06}, at least 20\% of the stars with [Fe/H]< -2.0 are CEMP stars. In these stars, the distribution of the C abundance is bimodally centred around $\rm A(C) \simeq 7.0$ (low-C band) and $\rm A(C) \simeq 8.3$ (high-C band) \citep{BonifacioCS15,bonifacio18}. With A(C)\,=\,6.9, HE\,1413-1954 belongs to the low-carbon band \citep[see fig. 2 in][]{bonifacio18}.

\subsection{$\alpha$ elements magnesium, silicon, calcium, and titanium}

\subsubsection{Magnesium}
Magnesium is produced during hydrostatic phases, specifically, in the burning of carbon in the shells of massive stars.
The Mg abundance from the three Mg lines blended by hydrogen lines (around 383\,nm) was deduced by spectrum synthesis.
The stars of our sample show positive and fairly constant ratios of [Mg/Fe] with values of $\rm \langle [Mg/Fe]\rangle = +0.24 \pm 0.10 \,dex$.

The CEMP star HE\,1413-1954 has the highest [Mg/Fe] ratio of our sample. This has been observed  for several CEMP stars.  
The star CS\,30302-145 exhibits a significantly low Mg value,  \\
$[{\rm Mg/Fe}]=-0.21$, presenting a difference of 0.42\,dex from the average value. This star was not taken into account in the computation of the mean [Mg/Fe] ratio. In Figs. \ref{fig:alpha}, \ref{fig:odd}, \ref{fig:ironpeak}, and \ref{fig:heavy}, the position of this star is indicated by a blue square.

\subsubsection{Silicon} \label{sec:si}
Silicon is formed during the oxygen burning and in smaller amounts, also by neon burning, in a combination of pre-explosive and explosive phases in type\,II supernovae (SNe). 
Abundances of silicon were determined using the line 390.55\,nm.
The mean value of [Si/Fe] is close to zero. The average value is $\rm 0.00 \pm 0.22$\,dex.
Following \citet{ShiGM09}, in the range of temperature and metallicity of the stars we studied, the LTE abundance of Si determined from the 390.55\,nm line should be corrected by about +0.25 dex. (This correction was not applied in Table \ref{tab:Abundance}).

With [Si/Fe]=--0.49 dex, the star CS\,30302-145, which was found to be Mg-poor, is also very Si-poor. (It was not taken into account in the computation of the mean [Si/Fe] ratio.)

\subsubsection{Calcium}

The Ca lines available in stellar spectra allow to derive A(Ca) in a wide range of metallicity.
We took advantage of the investigation of the departure from local thermodynamical equilibrium (NLTE) in \citet{spite12} to correct our measurements. The NLTE correction in our stars is about +0.1 dex. The values given in Table\,\ref{tab:Abundance} are corrected for NLTE effects. Moreover, we did not use the resonance line of neutral calcium at 421.6\,nm because in metal-poor stars, the use of this line leads to an underestimation of the Ca abundance, which reaches up to 0.3\,dex at [Ca/H]=--3.0
\citep[see e.g. ][]{spite12}. 
The mean abundance we found is $\langle [{\rm \ion{Ca}{i}/Fe}]\rangle =+0.46\pm 0.18$\,dex. 

The abundance of Ca deduced from the ionised lines is based on one single measured line, the \ion{Ca}{ii}-K line at 392.3\,nm. The Ca abundance derived from \ion{Ca}{i} and \ion{Ca}{ii} lines generally agrees well (see Table\,\ref{tab:Abundance}).

However, we note that CS\,30302-145 is an exception, and if we had used the abundance of Ca deduced from the \ion{Ca}{ii} line, this star, poor in Mg and Si, would have the lowest value of [Ca/Fe]  of our sample ([Ca/Fe]=\,0.2\,dex).

\subsubsection{Titanium}
The theoretical studies of core-collapse SNe and hypernovae predict the ejection of large quantities of titanium \citep{limongi03,chieffi03}. The explosion mechanisms have a substantial impact on the production of titanium.
The abundance of titanium was derived using the lines of ionised titanium, which is the dominant species in our sample of stars. The titanium-to-iron ratio is very consistent in all the stars of the sample: $\langle[{\rm Ti/Fe}]\rangle =\,0.38\pm 0.17$\,dex.

\subsection{Light odd-Z metals: sodium, aluminium}

\subsubsection{Sodium}
In massive stars, sodium is created mostly during hydrostatic carbon burning and partly during hydrogen burning through the NeNa cycle \citep[see e.g.][]{cristallo09,RomanoKT10}.
Here we studied the sodium D doublet lines (588.99\,nm and 589.59\,nm), which are the only sodium lines observed in EMP stars. These lines are affected by NLTE effects. In Table \ref{tab:Abundance} we applied the NLTE corrections provided by \citet{andrievsky07}.
In these warm metal-poor turnoff stars, the correction is only about $\rm-0.07\,dex$.
The mean value of [Na/Fe] in our sample is $\rm \langle[Na/Fe]\rangle =-0.43\pm 0.22$\,dex.\\
The Mg-Si-poor star CS\,30302-145  has the lowest sodium abundance ratio [Na/Fe]=--0.90.

\subsubsection{Aluminium}
In massive stars, aluminium is a product of carbon and neon burning \citep{limongi03,chieffi03,nomoto13b}. 
We analysed the  resonance doublet of \ion{Al}{i} at 394.40 and 396.15\,nm. 
Both lines are sensitive to departure from LTE. We applied the corrections by \citet{andrievsky08}. 
In these metal-poor turnoff stars, the NLTE correction is positive and very significant. It increases the LTE Al abundance by about +0.65\,dex. The values given in Table \ref{tab:Abundance} are corrected for NLTE. The mean [Al/Fe] value is --0.12$\pm$0.10\,dex (if we do not take into account CS\,30302--145).

The  [Al/Fe] value of the Mg-Si-poor star CS\,30302--145 is 0.55\,dex lower than the average value (see Fig.\,\ref{fig:odd}).

\subsection{Elements formed during complete and incomplete Si burning}

We included the iron peak elements Sc, Cr, Mn, Fe, Co, and Ni.
Iron is the most tightly bound nucleus, and the iron peak elements are the last for which fusion reactions are the main mode of production. 
In the matter that formed the old very metal-poor stars, these elements were ejected by type\,II SNe, and their production occurs during Si burning in the pre-SN phase and in the explosive phase \citep{limongi03,chieffi03}. 

\subsubsection{Scandium}
Scandium is exclusively observed in its singly ionised state. 
Our scandium abundances were measured using the \ion{Sc}{ii} line at 424.68\,nm. When comparing our results with the First Stars sample \citep{bonifacio09}, we have an excellent agreement.

The nucleosynthesis of this element is linked to the mass of the parent stars \citep{chieffi02}. 
If the gas out of which these stars were formed had been enriched by
a few SNe with slightly different masses,  we could expect a high dispersion in scandium abundances, 
which in our case (Fig.\,\ref{fig:odd}) is not observed. 

There is one exception: the very lithium-poor star CS\,22882--027. 
This star stands out with the lowest scandium abundance at about 3 $\sigma$ from the mean. 

The chemical evolution models of the Galaxy do not represent the evolution of the abundance of this element well \citep{kobayashi06,Matteucci16}. They generally underestimate the scandium abundance at low metallicity.
 
\subsubsection{Chromium}
The explosive combustion of  silicon  is the main source of chromium \citep[see e.g.][]{woosley95,chieffi02,umeda02}. 
Our LTE analysis based on \ion{Cr}{i} lines shows the well-known decrease in  [Cr/Fe] with decreasing metallicity. Following \citet{bergemann10}, this is an effect of NLTE on neutral chromium. The NLTE correction is about +0.4\,dex at [Fe/H]=--3 and only +0.2 \,dex at [Fe/H]=--2.
If we apply this correction to our sample of stars and to the stars studied in \citet{bonifacio09}, the ratio [Cr/Fe] is practically constant and close to zero in our metallicity range. The values given in Table \ref{tab:Abundance} are not corrected for NLTE. 

\subsubsection{Manganese}
At low metallicity, manganese is mainly made during silicon burning in core-collapse SNe. Our determination of the abundance of Mn in our sample of turnoff stars is based on the resonance Mn triplet at 403 nm, which is known to be strongly affected by departure from LTE \citep{BergemannGheren08,bergemann19}. This NLTE correction has not been applied in Table \ref{tab:Abundance} or in Fig \ref{fig:ironpeak}.
The average value is $\langle [{\rm Mn/Fe}]\rangle=-0.64$. 
Moreover, in Fig.\,\ref{fig:ironpeak},  the most metal-poor stars seem to be more Mn poor, but this can be explained by the strong increase in NLTE correction when the metallicity decreases (see Fig.\,9 in \citet{bergemann19}), but this correction has not been computed below [Fe/H]=--3.

Here again the Mg-Si-poor dwarf CS\,30302-145 stands out in the sample. It shows a considerably higher manganese abundance than the remaining sample (see Fig.\,\ref{fig:ironpeak}). This latter value has not been taken into account in the computation of the average (its value is 0.56\,dex above this average).

\subsubsection{Cobalt}
In our very metal-poor turnoff stars, the Co abundance is derived from the \Cou~ lines at 384.5, 399.5, and 412.1 nm. These lines are known to be widened by hyperfine splitting structure and are very sensitive to departure from LTE  \citep{bergemann08,bergemann10b}.
In our classical LTE analysis, we observe an increase in [Co/Fe] as the stellar metallicity decreases (see Fig.\,\ref{fig:ironpeak}), as was noted by~\citet{mcWilliam95}, \citet{cayrel04}, and \citet{bergemann10b}, for example.
According to \citet{bergemann10b}, the NLTE correction leads to an even stronger increase in [Co/Fe] with decreasing [Fe/H]. Chemical evolution models of the Galaxy fail to explain this behaviour \citep[see e.g. ][]{RomanoKT10,Matteucci16}.  \citet{bergemann10b} suggested that compared to the predicted yields, Co is overproduced relative to Fe in short-lived massive stars. 

In Fig.\,\ref{fig:ironpeak} the Mg- and Si-poor dwarf CS\,30302-145 has the highest [Co/Fe] ratio of our sample.

\subsubsection{Nickel}

The mean value of the nickel-to-iron ratio is about zero (Fig. \ref{fig:ironpeak}), $\rm\langle[Ni/Fe]\rangle=0.04 \pm 0.15$\,dex, in good agreement with the data of \citet{bonifacio09}.\\
The two stars CS\,30302-145 and HE\,1413-1954 have the highest values of the Ni abundance with respect to iron, see Fig.\,\ref{fig:ironpeak}. The error on the Ni abundance of the second star is very large, however.

\subsection{Neutron-capture elements Sr and Ba}

\begin{figure}
\centering
\resizebox{5.0cm}{!}
{\includegraphics{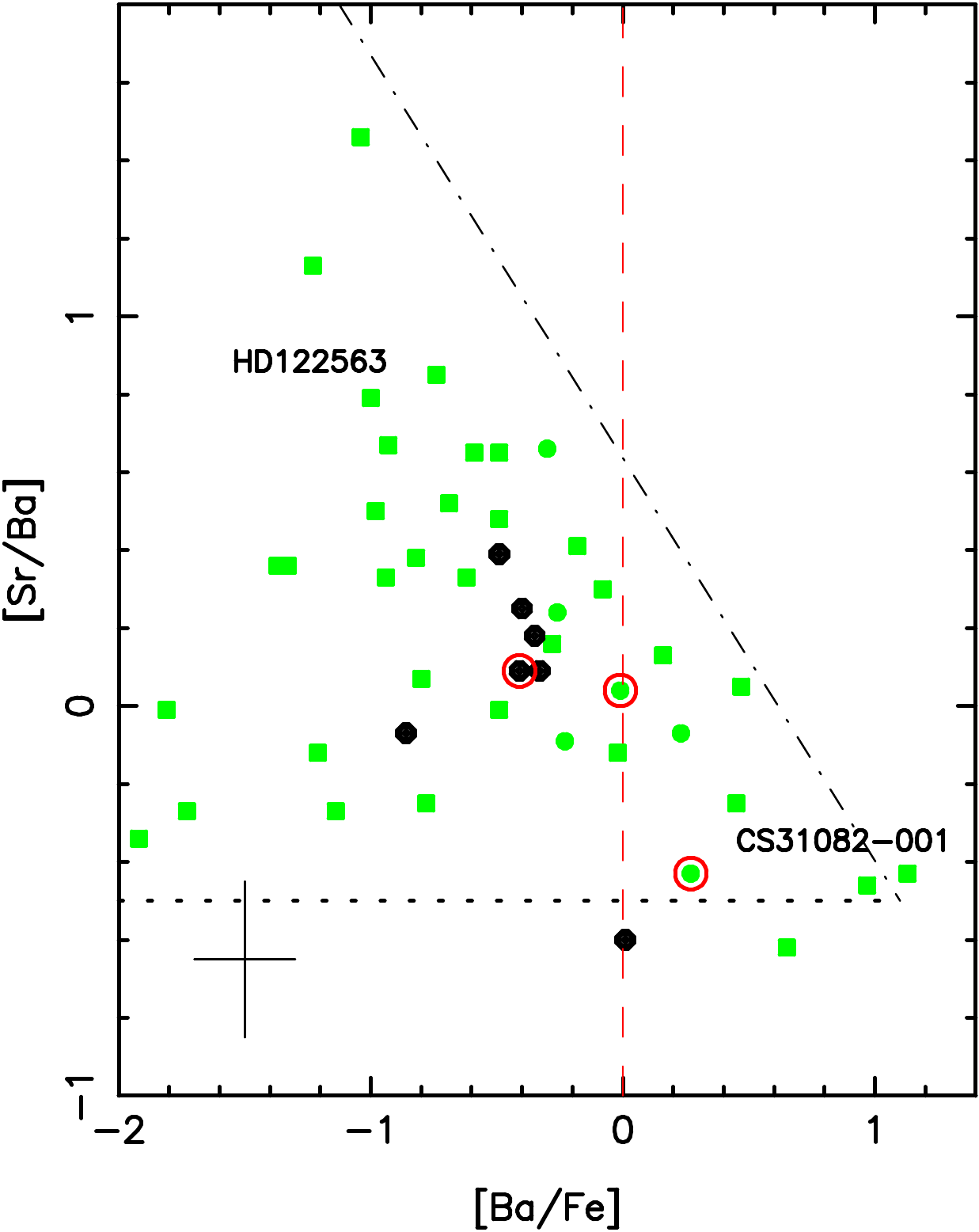}}
\caption{[Sr/Ba] vs. [Ba/Fe]. Green symbols represent the First Stars, circles show dwarfs, and squares show giants. The black dots represent our sample of dwarfs (Ba could be measured in only seven stars of the sample). The three Li-poor dwarf stars (with a measured Ba abundance) are circled in red. The dotted line represents the value [Sr/Ba] provided by the  $r-$ process alone.  CS\,31082-001 is a typical r-rich star, and in contrast, HD\,122563, is an r-poor star with a relative enrichment of the first peak elements.}
\label{fig:SrBa}
\end{figure}

While elements up to iron are formed by nuclear fusion in an exothermic way, for heavier nuclei, fusion reactions become endothermic because Fe nuclei are the most tightly bound. As a consequence, the preferred route for the formation of heavier nuclei is neutron capture \citep[see e.g.][and references therein]{ArnouldGor20,CowanSL21}.
This capture can occur through two main processes.
A very rapid process (main r-process) in violent events such as explosions following the core collapse of massive SNe, the merging of neutron stars or of black holes, jets, gamma ray bursts, etc. 
A slow process (main s-process) at a rate much lower than the $\beta$ decay. This occurs at the end of the evolution of relatively low-mass stars in their asymptotic giant branch (AGB) phase. 
Because these low-mass stars have a long lifetime, they could not enrich the matter that formed the very old metal-poor stars we studied. In our sample of stars, the abundance of the neutron-capture elements mainly reflects the products of the main r-process.\\ 
However, the large scatter of [Sr/Ba] for instance for a given [Ba/Fe]  suggests that another mechanism is able to enrich the matter in the early Galaxy \citep[e.g.][]{spite14,CowanSL21}. In Fig.\,\ref{fig:SrBa}, the ratio [Sr/Ba] is plotted versus [Ba/Fe] for the stars of the First Stars sample (dwarfs and giants) and our sample of stars. A star can be Sr rich but Ba poor. Contributions of fast-rotating massive stars through a non-standard s-process \citep{MeynetEM06,FrischknechtHT12,FrischknechtHP16}, or supermassive AGB through an intermediate i-process \citep{CowanRose77} are generally evoked. The contributions of these other processes become relatively high only in Ba-poor stars, where the matter that formed the star has been little enriched in heavy elements by the main $r-$process.

In warm very metal-poor dwarf stars such as those in our sample, only the abundances of Sr and Ba are measurable in the visible region of the spectrum. In this study, we analysed two ionised strontium lines at 407.7\,nm and 421.6\,nm and only the Ba line at  493.4\,nm because the strongest Ba line at 455.4\,nm is generally outside the wavelength range of our spectra: 
the Ba 455.4\,nm line could be measured only in BS\,17572-100  and HE\,1413-1954.

We could measure the Ba abundance in the C-rich star  HE\,1413-1954, and we found (Table \ref{tab:Abundance}): $\rm[Ba/Fe]=0.01 \pm 0.2$. This star, which belongs to the low-C band, is as expected not enriched in Ba \citep[see e.g. Fig. 6 in ][]{BonifacioCS15}. HE\,1413-1954 can be classified as a CEMP-no star \citep[following][]{BC05}.

\section{Kinematic and orbital properties of the star sample }

\begin{table*}
\caption{Galactic coordinates, velocities, and main orbital parameters of the stars we studied (the units are given in the text). We add the parameters of the three Li-poor stars studied in the frame of the First Stars project at the end of the table. }
\scalefont{0.72}
\begin{tabular}{lrrrrrrrrrrrrrrr}
\hline
    &            &        &        &           &         &        &          &    &       &     &       &       &     &         & arc tang \\
Star&           X&       Y&       Z&    $V_{X}$&  $V_{Y}$& $V_{Z}$&  $V_{R}$ &Rmax&  z max&   Lz&   Vphi&  Lperp&  ecc&  Energy &(zmax/ \\
    &            &        &        &           &         &        &          &    &       &     &       &       &     &         &   Rmax) \\
\hline 
\multicolumn{16}{l}{Our sample of turnoff stars with A(Li) compatible with the plateau}\\
BS17572-100&  -8.41&  -0.33&   0.25&    8.23&   -3.32&   19.18&   -8.09&  8.45&  3.84&   0.31&    3.64&  1.64&  0.98&  -1814& 0.43  \\
CS22950-173&  -7.63&   0.33&  -0.37&  281.12&   17.35&   80.71& -280.11& 17.16&  5.09&  -2.25&  -29.42&  5.12&  0.92&  -1438& 0.29\\
CS29491-084&  -7.67&   0.21&  -0.94&  43.40&   48.46&   -8.37&  -42.06&  7.90&  1.03&   -3.81&  -49.63&  1.14&  0.79&  -1833& 0.13\\
CS29514-007&  -8.30&   0.00&  -1.17& -207.90& -361.06&  -16.80&  208.00& 47.10&  6.42&  29.95&  361.00&  4.34&  0.75&   -940& 0.14\\
CS29516-028&  -8.02&   0.53&  -0.50& -86.78&  127.08&  114.55&   95.04&  9.35&  4.63&   -9.72& -121.03&  9.70&  0.45&  -1654& 0.46\\
CS30302-145&  -7.18&  -0.18&  -0.54&   72.18& -300.69& -113.72&  -64.2&  16.86&  4.89&  21.70&  302.45&  8.67&  0.42&  -1354& 0.28\\
CS30344-070&  -7.55&   0.09&  -1.27&  -8.43&  128.34& -168.80&    9.92&  8.19&  5.90&   -9.68& -128.23& 12.72&  0.19&  -1630& 0.62\\
HE1413-1954&  -6.97&  -0.75&   1.16& -175.35&  164.93&   -7.33&  156.63& 10.71&  2.03& -12.82& -182.81&  3.15&  0.48&  -1604& 0.19\\
HE0148-2611&  -8.45&  -0.15&  -1.17&  132.99&  -66.46&  254.10& -131.82& 17.13& 16.22&   5.81&   68.74& 19.96&  0.51&  -1367& 0.76\\
\\
\multicolumn{16}{l}{Our sample of turnoff stars  with A(Li) significantly lower than the plateau}\\
CS22188-033&  -8.16&  -0.07&  -0.43& -53.61&   -9.28&   28.02&   53.69&  8.44&  0.74&   0.72&    8.81&  2.52&  0.95&  -1812& 0.09\\
CS22882-027&  -8.10&  -0.06&  -1.31& -40.62& -160.79& -162.76&   41.73&  9.95&  6.21&  13.00&  160.51& 12.81&  0.22&  -1549& 0.56\\
\hline\hline
\\
\multicolumn{16}{l}{Turnoff stars of the First Stars sample with A(Li) significantly lower than the plateau}\\
CS22888-031&  -7.85&   0.03&  -0.85&  -215.05&    43.68&   40.58&  215.23& 12.54&  4.70& -3.36&  -4.27&   5.02& 0.89&   -1595&  0.36\\
CS22948-093&  -6.87&   0.01&  -1.62&   397.50&    97.11& -147.46& -397.42& 48.31& 42.60& -6.69&  -9.74&  16.65& 0.88&    -944&  0.72\\
CS22966-011&  -7.96&   0.08&  -0.84&   -42.80&    27.84&  -14.21&   43.07&  8.23&  3.73& -2.18&  -2.74&   0.80& 0.89&   -1823&  0.43\\
\hline
\multicolumn{16}{l}{The orbital characteristics of the stars studied in the First Stars sample are given in \citet{DiMatteoSH20}.}\\
\end{tabular}
\label{tab:kine}
\end{table*}

In this section, we compare the kinematic and orbital parameters of our 11 stars to the First Stars sample that was extensively studied in \citet{DiMatteoSH20}. In particular, we try to detect, for instance, whether the Li-poor stars (from our sample or the  First Stars sample) present any special characteristics.\\
We used the GalPot code\footnote{\url{https://github.com/PaulMcMillan-Astro/GalPot}} \citep{DB98} together 
with the galactic model of \citet{mcmillan17} to derive the kinematical properties of our sample of stars. In this model, the distance of the Sun from the Galactic centre is $ \rm R_{\odot} = 8.21\,kpc$ and the local standard of rest (LSR) velocity, $ \rm V_{LSR} = 233.1$ \kms \citep{mcmillan17}. The velocity of the Sun relative to the LSR is $U_{\odot}=11.1$\,\kms, $V_{\odot}=12.24$\,\kms~ and $W_{\odot}=7.25$\,\kms \citep{mcmillan17}. The disc rotates clockwise, and as a consequence, the z-component of the disc angular momentum, Lz, and the disc azimuthal velocity, Vphi, are negative. A star with negative Lz and Vphi is prograde.

The positions, proper motions, and parallaxes of our sample of stars are taken from Gaia DR2, but the
radial velocities were measured on the spectra (Table 1 in Paper I) with a precision of 1\,\kms. 
The sample of stars we studied has about the same distance distribution, between 0.4 and 1.9 kpc (Table \ref{tab:ParametrosEstelares}), as the sample of dwarfs studied in the frame of the First Stars \citep{DiMatteoSH20}, which is used here also as a chemical comparison sample. The error on the parallax for all the stars is lower than 12\%  (see Table \ref{tab:ParametrosEstelares}). The main kinematical data and orbital properties of these stars are presented in Table \ref{tab:kine}. 

\noindent$\bullet$ X, Y, Z are the galactic coordinates of the stars, and $ V_{X},  ~V_{Y}, ~V_{Z }$ their velocities in the three directions. X, Y, Z, are given in kpc and the velocities in \kms.\\
$\bullet$ $R=\sqrt{X^{2} + Y^{2}}$ is the distance of the star to the Galactic centre  in kpc, and $ V_{R} $ is the component of the velocity in this direction:  $ V_{R} = (X V_{X} + Y V_{Y})/R$.\\
$\bullet$ Lz, the z component of the angular momentum, is equal to $X V_{Y} - Y V_{X}$.\\ 
$\bullet$ Lperp, the perpendicular angular momentum component, is equal to $\sqrt{Lx^{2}+Ly^{2}}$ (with $Lx= Y V_{Z} - Z V_{Y}$ and $Ly= Z V_{X} - X V_{Z}$).
Angular momenta are given in units of 100 kpc \kms.\\
$\bullet$ The azimuthal velocity Vphi is defined as the z component of the angular momentum divided by R,  that is, ~$\rm Vphi = Lz / R$.\\
$\bullet$ The energy of the orbit is given in units of 100 km$^2$\,s$^{-2}$.\\

\begin{figure*}
\begin{center}
\resizebox{5.8cm}{!}
{\includegraphics [clip=true]{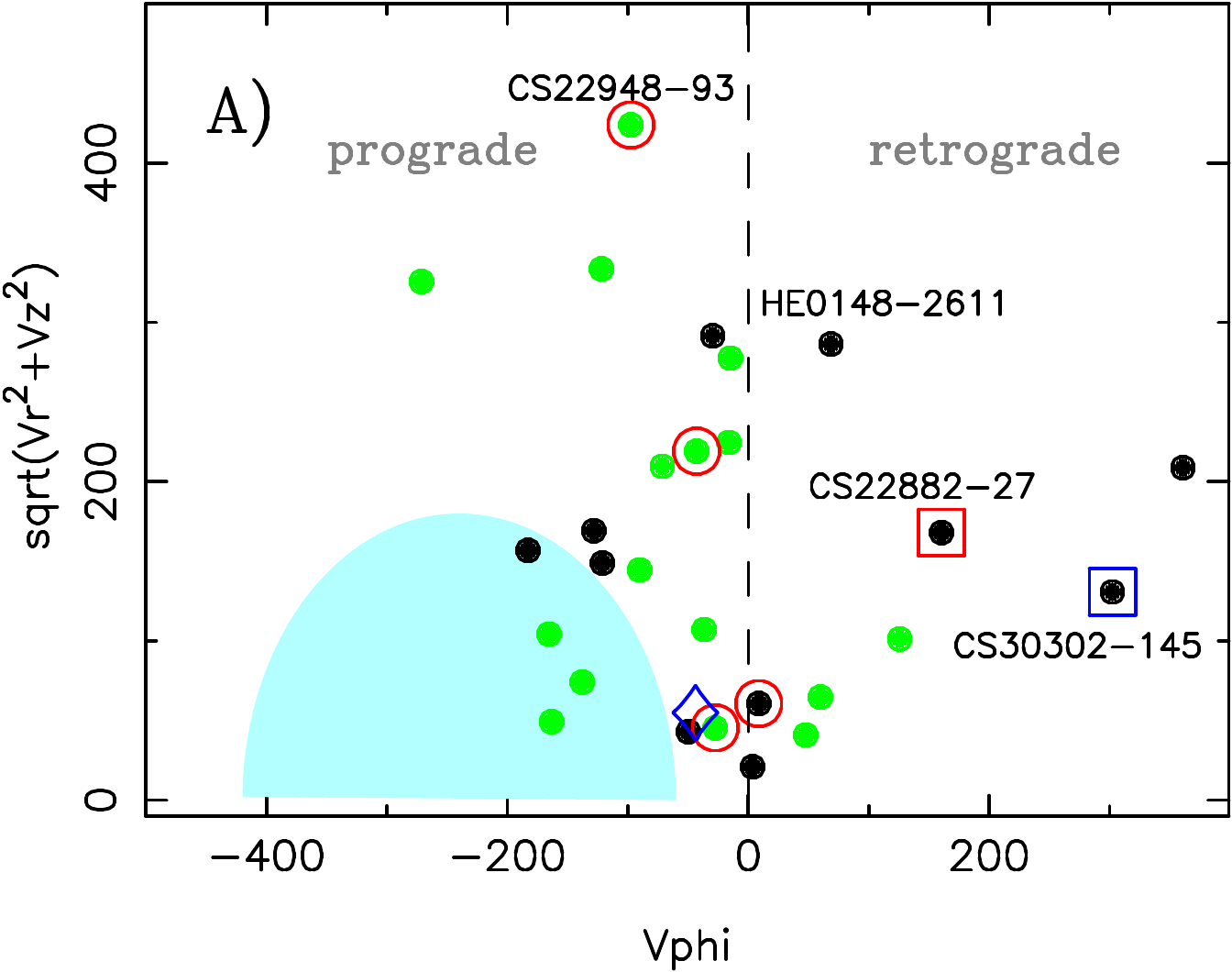}}
\resizebox{5.8cm}{!}
{\includegraphics [clip=true]{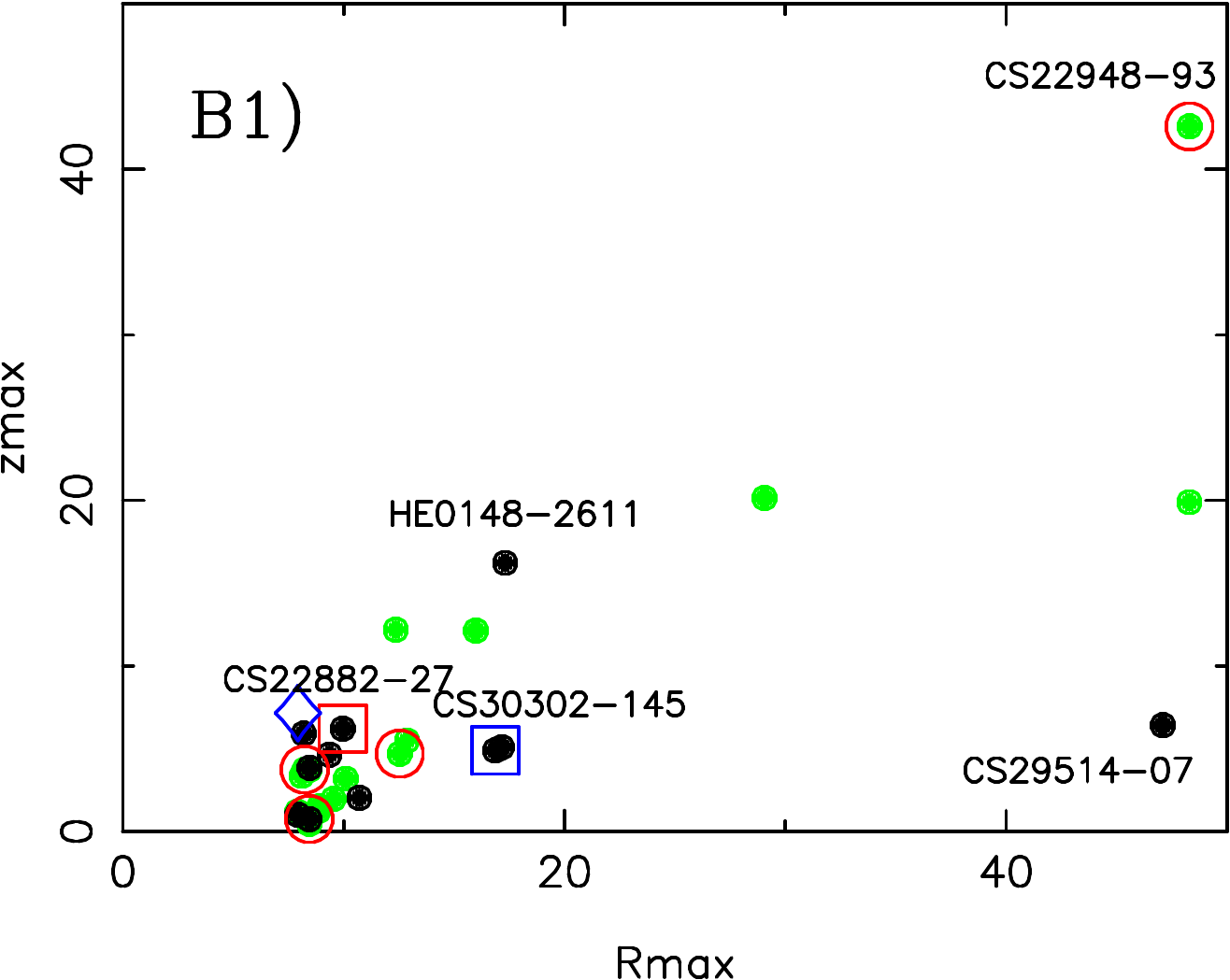}}
\resizebox{5.8cm}{!}
{\includegraphics [clip=true]{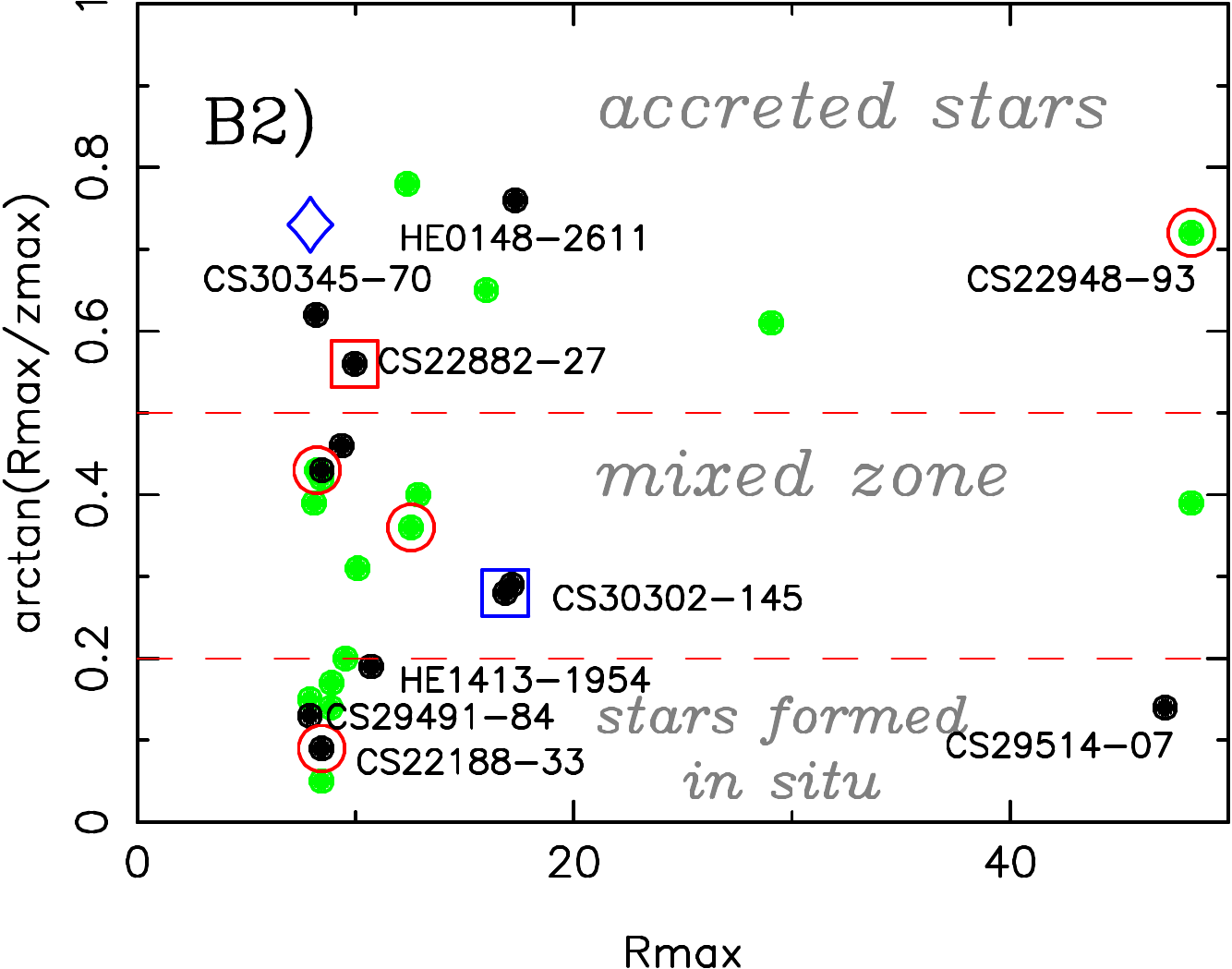}}
\resizebox{5.8cm}{!}
{\includegraphics [clip=true]{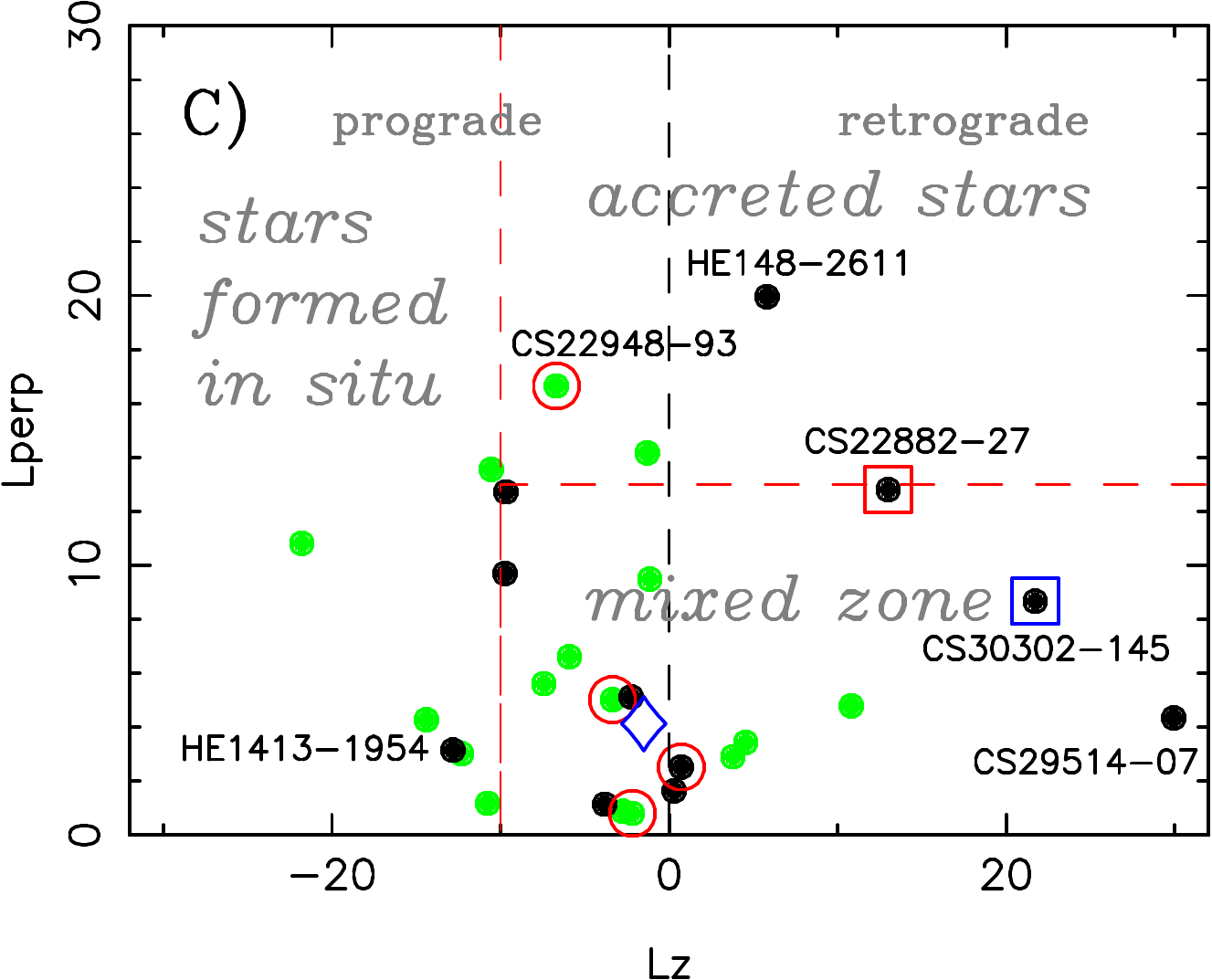}}
\resizebox{5.8cm}{!}
{\includegraphics [clip=true]{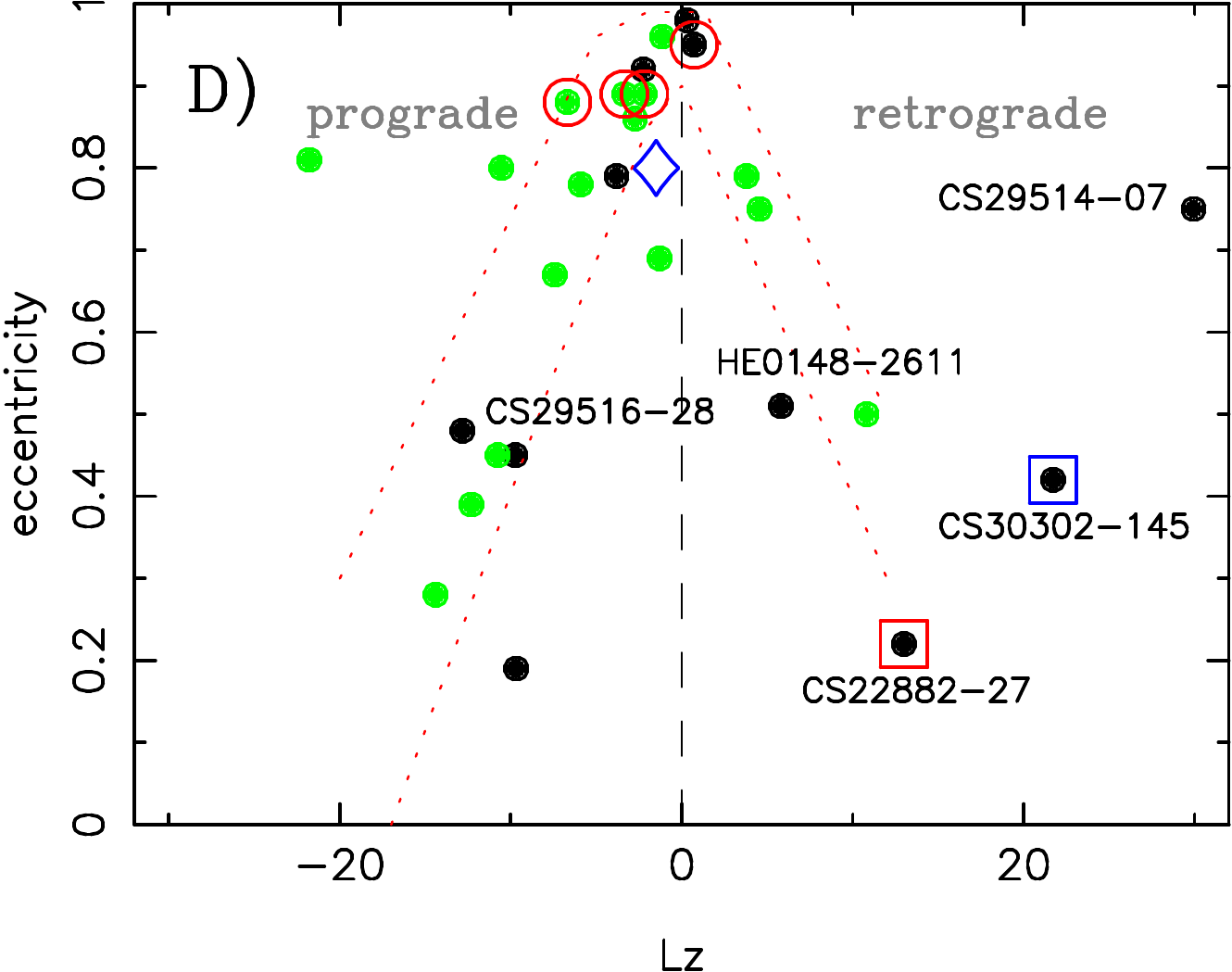}}
\resizebox{5.8cm}{!}
{\includegraphics [clip=true]{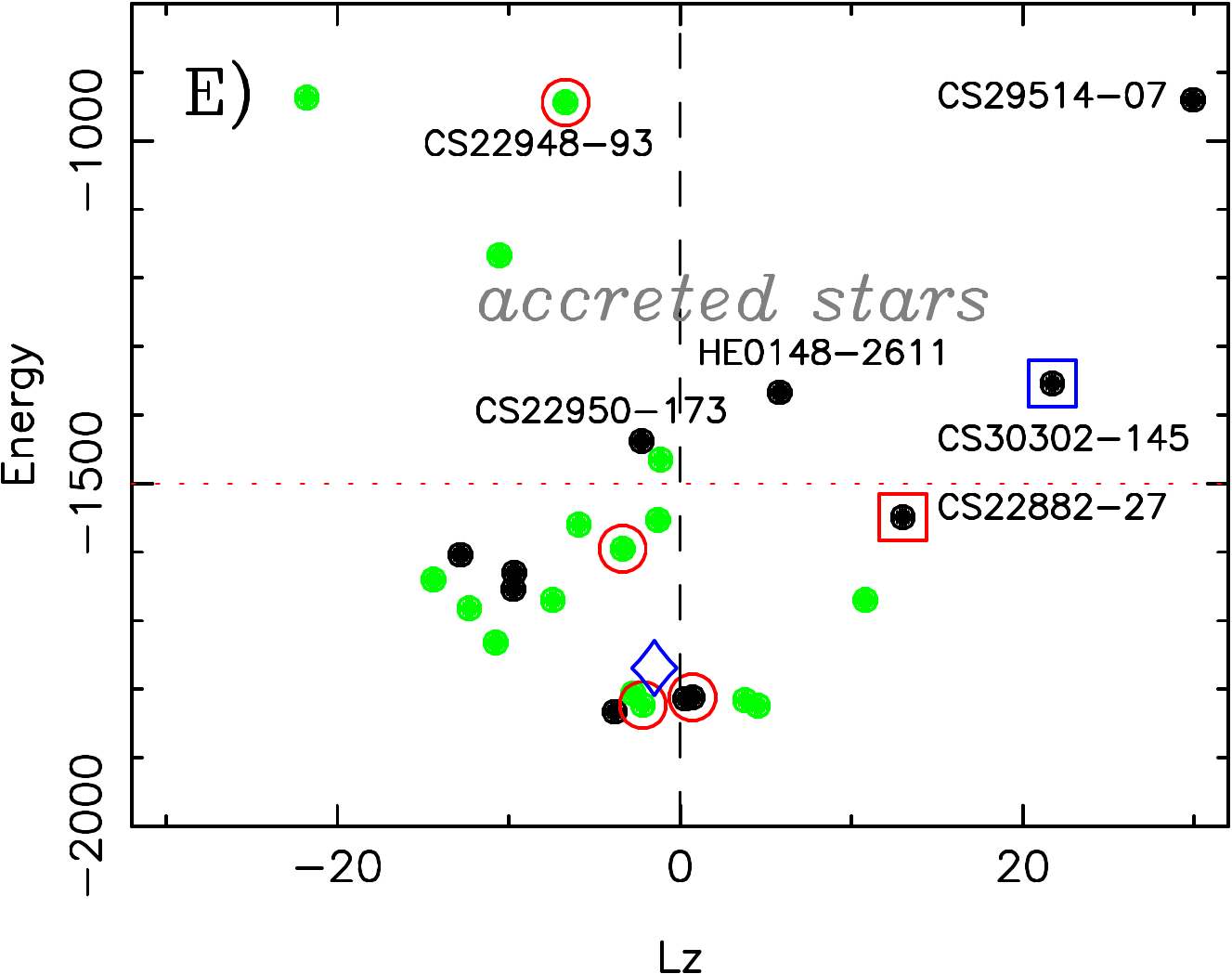}}
\end{center}
\caption{Characteristics of the orbital kinematics. Our stars  are represented by black dots, and the dwarfs of the First Stars by green dots. The Li-poor stars are circles in red, the Li-free star by a red square, and the low-$\alpha$ star by a blue square. The other very metal-poor and low-$\alpha$ star HE\,1424-0241 \citep{cohen07} is indicated by an open blue diamond. In panel A, the blue zone shows the location of the disc stars in this diagram. In panels C and E, the location occupied mainly by accreted stars is indicated. In panel D, the dotted lines delimit the region in which the \citet{nissen10} stars are located.
Units of the plotted quantities are velocities in \kms, distances  in kpc, angular momenta in 100\,kpc\,\kms, and energy in 100 km$^2$\,s$^{-2}$.  It is important to note that in the zone marked as containing ``accreted stars", e.g., there is only a high probability that the stars in this zone are accreted, and this is the same for the zone marked ``in situ''. Exceptions may exist \citep{SchusterMN12}.  }
\label{fig:orbit}
\end{figure*}

In  Fig. \ref{fig:orbit} we compare the position of our stars (black dots) to the position of the First Stars (grey dots) in the five classical diagrams:\\ 
.~~~Fig. \ref{fig:orbit}a Toomre diagram,  Vphi vs. $ \rm \sqrt{V_{R}^{2} + V_{Z}^{2}}$;\\  
.~~~Fig. \ref{fig:orbit}b zmax vs. Rmax\\  
.~~~Fig. \ref{fig:orbit}c Lperp vs. Lz\\  
.~~~Fig. \ref{fig:orbit}d eccentricity vs. Lz\\ 
.~~~Fig. \ref{fig:orbit}e Energy vs. Lz. \\

Generally speaking, in Fig. \ref{fig:orbit} our new set of stars (black dots) occupies the same location as the dwarfs of the First Stars sample (green dots). The Li-poor stars (with A(Li) < 2.0) in our new star sample and in the First Stars sample are circled in red, the Li-free star CS\,22882-27 by a red square, and the low-$\alpha$ star CS\,30302-145 by a blue square. In the remainder of this paper, we discuss the position of our new sample of stars and of the Li-poor stars from our sample or the First Stars sample.

\citet{HaywoodDL18} and \citet{DiMatteoSH20} showed that the low-$\alpha$ and high-$\alpha$ stars of the \citet{nissen10} sample and the Gaia DR2-APOGEE sample \citep{DiMatteoHL19} occupy different positions in the diagrams of Fig.\,\ref{fig:orbit} \citep[see in particular Fig. 5 in ][]{DiMatteoSH20}. From these positions, it has been possible to define different zones populated mainly by accreted stars or stars formed in situ. These different zones are reported in Fig. \ref{fig:orbit}. 
Simulations suggested indeed that the low-$\alpha$ stars in the \citet{nissen10} and the Gaia DR2-APOGEE samples are the remnants of an accretion event in the early history of the Milky Way. The stars in these samples are more metal rich than the stars of the First Stars sample and the 11 stars we studied, but \citet{DiMatteoSH20} showed that despite their different abundance content, they can share the same common origin. \\
 However, we must insist on the fact that there is an important overlap in kinematics properties of the stars. It clearly appears that there is no sharp transition between the accreted stars and the stars formed in situ (in contrast to what might be thought from our   Fig. \ref{fig:orbit}). A star in a region noted ``accreted stars'' in  Fig. \ref{fig:orbit} has only a higher probability to be an accreted star than a star located in a region noted ``stars formed in situ''.

The Toomre diagram (see Fig. \ref{fig:orbit}A) shows that, like the First Stars sample, our stars are mainly prograde. The light blue area shows the location of the disc stars. Most of our stars are rather close to this zone. 
The azimuthal velocity Vphi (and the z component of the angular momentum Lz) of the Li-poor stars is generally close to zero.

Most of the 11 stars we studied remain during their life in or close to the thick disc  with $ \rm zmax \leq 5\,kpc$, (see Fig. \ref{fig:orbit}\,B1).  Only HE\,0148-2611  and the Li-poor star of the  First Stars sample, CS\,22948-093, have a  zmax value higher than 10\,kpc.  Two stars with Rmax > 40\,kpc are moving far away from the Galactic centre : CS\,29514-07 and the Li-poor star of the  First Stars sample, CS\,22948-93.\\ 
Following \citet{DiMatteoSH20}, the  stars with $\rm arctan(Rmax/zmax) < 0.2$  (see Fig. \ref{fig:orbit}\,B2)  are in a zone in which most of the stars are formed in situ. In the zone where  arctan(Rmax/zmax) values are above 0.5, the stars are in contrast probably accreted. 
(The limits between accreted stars and stars formed in situ are discussed in detail in section \ref{acc}).\\
One Li-poor star CS22188-33 is in the zone of the ``in situ stars''  , but another Li-poor star, CS22948-93, is in the zone of the ``accreted stars''. The Li-free star CS22882-27 is in the zone of the ``accreted stars''.

In Fig. \ref{fig:orbit}\,C, only one star of our sample (HE\,1413-1954) is in the region Lz < --10\,kpc\,\kms  , in which following \citet{DiMatteoSH20}, most of the stars were formed in situ.\\ 
Two stars with Lz > --10\,kpc\,\kms~ associated with $ \rm Lperp \gtrsim$ 13\,kpc\,\kms, are in a zone only populated by accreted stars, following \citet{DiMatteoSH20}:    HE\,0148-2611 and the Li-poor star CS\,22948-93.  The Li-free star CS22882-27 is at the limit of this zone, which is mostly occupied by the accreted stars.\\
The region with of Lz > --10\,kpc\,\kms \ but Lperp < 13\,kpc\,\kms \ is the locus in which stars of the Gaia-Sausage-Enceladus structure \citep[GSE, ][]{belokurov18,HaywoodDL18,helmi18} are redistributed, but also the locus of in-situ stars is heated to halo-like kinematics by accretion \citep{DiMatteoHL19}. 
The $\alpha$-poor star CS\,30302-145 belongs to this mixed region.

In Fig. \ref{fig:orbit}\,D, the relation between the eccentricity and the angular momentum Lz is presented. The locations of most of the Gaia-DR2-APOGEE and  \citet{nissen10} stars are indicated by dotted lines. The four Li-poor stars have an angular momentum close to zero, and as most of the stars with this Lz value, they have a high eccentricity (ecc > 0.85) \citep[see][]{DiMatteoSH20}. 

Fig. \ref{fig:orbit}\,E shows the relation between the energy E and the angular momentum Lz. Following \citet{DiMatteoSH20}, stars with E > --1500 $\rm km^{2}\,s^{-2}$ are most probably accreted because they lie in a region that is populated only by accreted stars in the Gaia-DR2-APOGEE and  \citet{nissen10} samples.  In this region, we find three stars of our sample, CS22950-173, HE014826-11, and the $\alpha$-poor star CS\,30302-145. The Li-poor star CS22948-93 also belongs to this ``accreted zone''.

The positions of our sample of stars in the different diagrams of Fig. \ref{fig:orbit} are summarised in Table \ref{tab:accret}. The Li-free CS22882-027 presents many characteristics of the accreted stars and also the Li-poor star CS22948-093. In contrast, the Li-poor star CS22188-033, with its very low value of arc tang (zmax/Rmax), seems to have been formed in situ. The other Li-poor stars CS22888-031 and CS22966-011 are in zones in which accreted stars and stars formed in situ are found (mixed zones).\\
 We note that the star CS29514-007, with its low value of $ \rm \arctan(Rmax/zmax),$ seems to have been formed in situ. In contrast, its very high energy value suggests that it has been accreted.

\section{Discussion}

\begin{table}
\caption{Status of the stars of our sample.}
\label{tab:accret}
\scalefont{0.9}
\begin{tabular}{lcccccrrrrrrrrrrr}
\hline 
           & arc tang           &  Lperp  &   Energy \\
           & (zmax/Rmax)        &  Lz     \\
           \hline
\multicolumn{6}{c}{stars with A(Li) compatible with the plateau}\\
BS17572-100&  \\
CS22950-173&                    &         &     acc \\
CS29491-084&     sit            \\
CS29514-007&     sit            &         &     acc \\
CS29519-028&     acc            &         &     acc \\
CS30302-145&                    &         &     acc \\
CS30344-070&     acc            &  acc    &\\
HE1413-1954&     sit            &  sit    &\\
HE0148-2611&                    &  acc    &     acc \\
\hline
\multicolumn{6}{c}{Li-free star }\\
CS22882-027&   acc     &  acc   &    (acc) \\
\hline
\multicolumn{6}{c}{Li-poor star }\\
CS22188-033&   sit\\
\hline\hline
\multicolumn{6}{c}{Li-poor stars  in the   First Stars}\\
CS22888-031&  \\
CS22948-093&   acc     &  acc   &     acc  \\
CS22966-011&  \\
\hline
\end{tabular}

The status (``sit'' is for ``formed in situ'' and ``acc'' for ``accreted'') is from the value of  arc tang (zmax/Rmax), the position of the star in the diagram Lperp vs. Lz and the value of the energy.
We added three Li-poor stars from the   First Stars sample at the end of the table. (The orbital characteristics of the stars studied in the First Stars sample are given in \citet{DiMatteoSH20}).
\end{table}


\subsection{Colour-magnitude diagram, age of the stars, and the blue straggler CS\,22882-027.}  \label{sec:age}
\begin{figure}
\centering
{\includegraphics[clip=true]{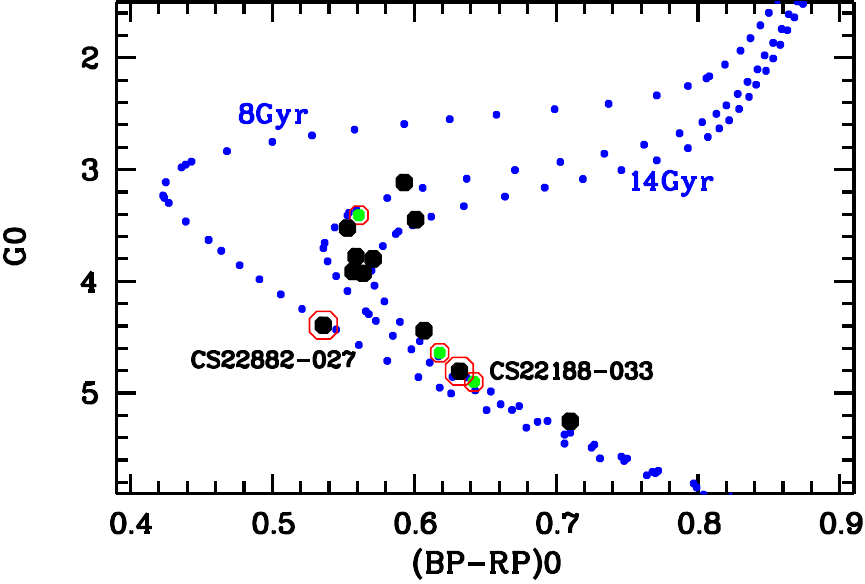}}
\caption{Comparison of PARSEC metallicity isochrones of --3.0 and ages 8, 12, and 14\,Gyr with the position of the stars in the colour-magnitude diagram. The colours $\rm(Bp-Rp)_{0}$ and the $\rm G_{0}$ magnitudes of the stars have been computed from the Gaia-DR2 photometry. The stars of our sample are represented by black dots, and the Li-poor stars (stars that deviate by more than 2 $\sigma$~from the Li plateau) are circled in red.
The three Li-poor stars from the  First Stars sample (green dots) have been added.}
\label{fig:isoparb}
\end{figure}

In Fig.\,\ref{fig:isoparb} we plot the position of the PARSEC metallicity isochrones --3.0 for different ages  (see also Sec. \ref{sec:stelpar}) in a $\rm G_{0}$ versus $\rm(Bp-Rp)_{0}$ diagram. The positions of the stars of our sample are marked by black dots, and the two Li-poor stars are circles in red. 

All the stars in Fig.\,\ref{fig:isoparb} are compatible with an age in the range of 12 to 14 Gyr, as expected from their low metallicity. \\
However, one star,  CS\,22882-027, seems to be much younger. 
According to the isochrone, its age is 8\,Gyr and its mass is 0.8 M\sun, slightly higher than the masses estimated for the other stars, which are in the range 0.67-0.78\,${\rm M}_\odot$.
Moreover, we have seen that it has no measurable lithium (Sec. \ref{sec:liab}).
These are two characteristics of blue stragglers \citep{bonifacio19}. A blue straggler is the result of the merging of two stars. Lithium is destroyed in the merging process, and the newly formed star, which has a higher mass, occupies a position on the colour-magnitude diagram that would normally be populated by genuinely younger stars. \citet{ryan01} suggested that the stars with no detectable Li are blue-stragglers-to-be  and \citet{bonifacio19} showed that three out of four of the stars investigated by \citet{ryan01} are indeed canonical blue-stragglers. We note that CS\,22882-027 has a normal abundance of all the other elements except for Sc,  whose abundance is very low (see Fig. \ref{fig:ironpeak}).

\subsection{Li-poor stars } 
In addition to the blue straggler  CS\,22882-027, another star of our sample, CS\,22188-033, has a rather low Li abundance, and during the Large Programme First Stars, three other stars were also found with a  Li abundance deviating by more than 2 $\sigma$ from the plateau  (see section \ref{sec:liab}). These three stars are plotted in Fig. \ref{fig:isoparb} as green dots.
All these lithium-poor stars have a normal distribution of the other elements. They do not present any systematic chemical anomaly.

It is interesting to remark that although we mainly studied stars located at the turnoff, three out of four Li-poor stars (excluding the blue straggler CS\,22882-027) are on the main sequence before the turnoff (see Fig. \ref{fig:isoparb}). This could suggest that in the most metal-poor dwarfs, some kind of atmospheric phenomenon (possibly diffusion) brings Li in the deep atmosphere, from where it is again mixed in the photosphere as the star reaches the turnoff and the surface convective zone deepens (see also Mucciarelli et al. to be submitted, private communication).

\subsection{Stars: accreted or formed in situ ? }  \label{acc}
The kinematic properties reflect large intrinsic differences between the stars formed in situ and those derived from accretion events. The most massive galaxies have a higher percentage of stellar accretion, and they harbour the most metal-poor stars \citep{oser10,Simon19}. 
Table\,\ref{tab:accret} provides the kinematical characteristics of the stars and their probability to be accreted or formed in situ. 

However, we must note that all our interpretation is based on the assumption that the high-$\alpha$ stars observed by \citet{nissen10} (with $\rm [Fe/H] \approx -1.2$) mostly have an in-situ origin, and that low-$\alpha$ stars are mostly associated with accretion events  \citep{DiMatteoSH20}. On the one hand, \citet{RenaudAR21} recently showed that a population of high-$\alpha$ stars can also be formed through accretion of massive satellites (see their Fig 8), and on the other hand, \citet{KhoperskovHS21} showed that a population of low-metallicity, low-$\alpha$ stars can also be formed in situ in the outer regions of Milky Way-type galaxies, where the star formation efficiency may be low (see some of the cases shown in Fig 8 of their paper). 

On the basis of their orbital properties, and if we assume that our interpretation of the observations of \citet{nissen10} is correct, we can deduce that  six stars of our sample have probably been accreted and that three stars were probably formed in situ (Table.\,\ref{tab:accret}).\\
The star  CS\,29514--007 presents ambiguous characteristics. It has a low value of arctan(Rmax/zmax), but a high Rmax (Fig. \ref{fig:orbit}). It lies in a region where both accreted and in-situ stars are found. It may be possibly accreted, as suggested also by its high energy value. This interpretation  does not contradict the finding of \citet{AmaranteSB20} and \citet{KoppelmanBH20}, who showed that structures or wedges discussed by 
\citet{HaywoodDL18} in the zmax-Rmax plane depend on the different orbital families populated by halo stars and resonant orbits.

\subsection{$\alpha$-poor star CS\,30302-145}

The EMP star CS\,30302-145 we studied has a  low abundance of the $\alpha$ elements Mg and Si, but the [Ca/Fe] ratio is consistent with the sample average (see Fig. \ref{fig:alpha}). In this star, the odd light metals Na and Al (see Fig. \ref{fig:odd}) are also deficient, but it is clearly Mn-rich and belongs to the stars with the highest ratios of [Co/Fe]  and [Ni/Fe] (see Fig. \ref{fig:ironpeak}). 

\begin{figure}
\centering
\resizebox{8.1cm}{!}
{\includegraphics{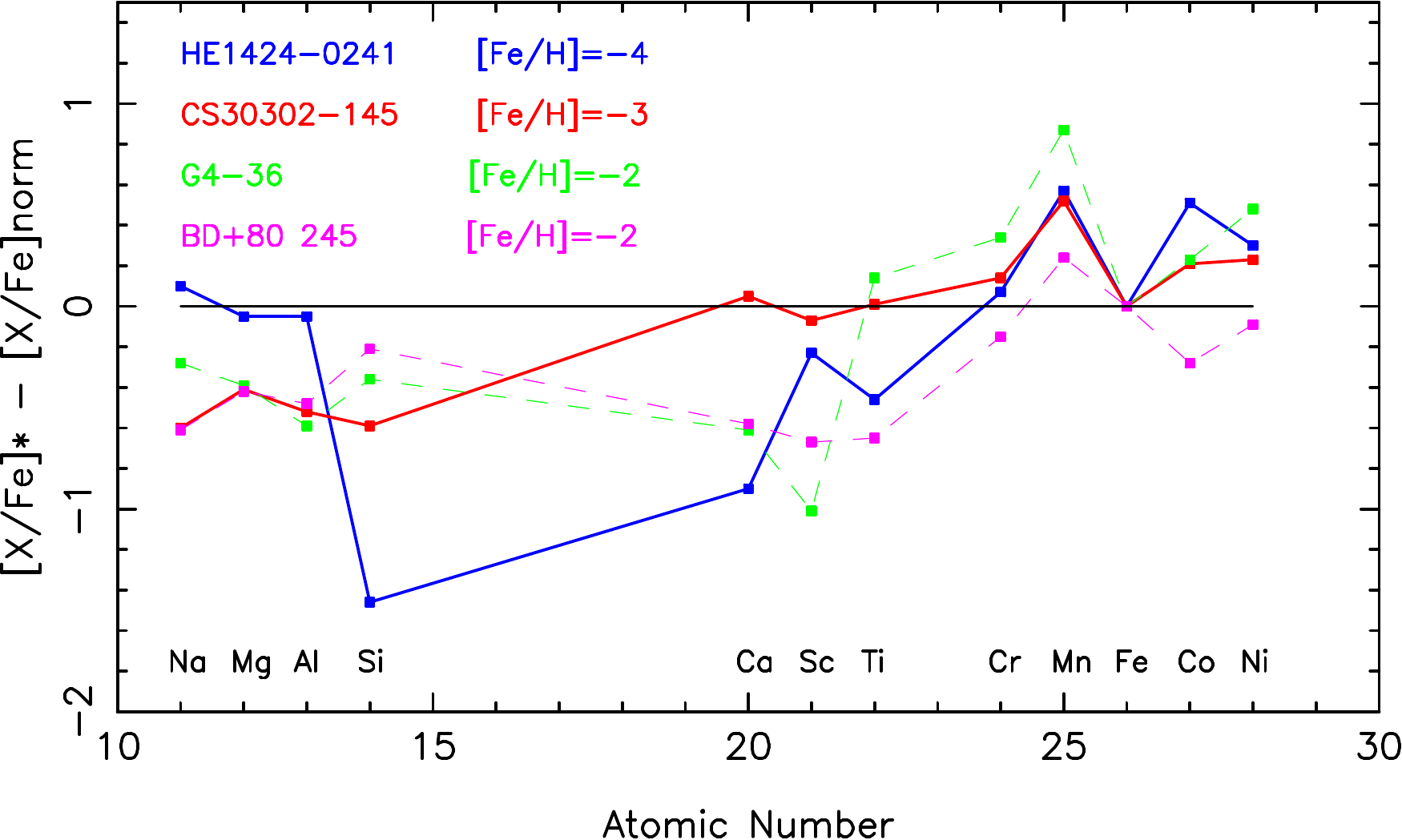}}
\caption{Comparison of the abundances of the $\alpha$-poor star. The star CS\,30302-145 in our sample (red line, whose metallicity is [Fe/H]$\sim$--3) compared with the stars in the literature, HE\,1424-0241 (blue line, with [Fe/H]$\sim$--4 \citep{cohen02}). Two less metal-poor but $\alpha$-poor stars BD+80 245 (pink line), and G4-36 (green line) have been added to this figure ([Fe/H]$\sim$--2 following \citet{ivans03}).}
\label{fig:pat-alfa-poor}
\end{figure}

A similar behaviour has been observed in another EMP star, HE\,1424-0241, a giant  with $\rm[Fe/H]\simeq -4$ \citep{cohen07}. This star also has a very low [Si/Fe] ratio, but in contrast to CS\,30302-145, Mg is about normal and Ca is deficient. Both stars are Mn rich. For these two stars, we computed the differences between their value of [X/Fe] and the value of [X/Fe] in similar ``normal'' stars. For  HE\,1424-0241, we took the normal giants from the 0Z survey as a reference, evaluated at [Fe/H]=--4 \citep[][column 7]{cohen07}, and for  CS\,30302-145, we adopted the mean value, estimated at [Fe/H]=--3, from \citet{bonifacio09}.
In Fig. \ref{fig:pat-alfa-poor} we plot this difference as a function of the atomic number. CS\,30302-145 is much less extreme than HE\,1424-0241 .

In Fig. \ref{fig:pat-alfa-poor} we also add (dotted lines) two less metal-poor  but $\alpha$-poor subgiants BD+80\,245 and G4-36 \citep{ivans03,SalvadoriBC19} with $\rm[Fe/H]\simeq -2$. For these stars, we adopted as a reference the mean value of [X/Fe] at [Fe/H]=--2.0 following \citet{reggiani17}.
Another $\alpha$-poor star (and r-process-enhanced star) has been reported by \citet{SakariRP19}, this horizontal-branch star with [Fe/H]=--1.9, RAVE J\,0937-0626, is also poor in Mg, Al, Si, and Ca, but it does not seem  to be enriched in Mn or in Co.
                                                         
All the stars in Fig. \ref{fig:pat-alfa-poor} present a similar trend, but the differences from star to star are significant. 
In BD+80\,245 and G4-36, which are less metal-poor ([Fe/H]=--2), \citet{ivans03}  tried to explain the distribution of the elements as caused by the effect of the first SN\,Ia. They concluded that the abundance distribution they observed could only be explained by a strong contribution of the SN\,Ia events. 
The same explanation was suggested by \citet{SakariRP19} for RAVE J\,0937-0626.\\ 
This is unlikely to occur in stars as metal-poor as CS\,30302-145 or HE\,1424-0241. The matter that formed these two stars could be enriched only by the ejecta of SN\,II, and following \citet{cohen07}, the different models of SN\,II yields are not able to explain the peculiar distribution of the elements.

\citet{SalvadoriBC19} suggested that BD+80\,245 ([Fe/H]=--2.0) had been enriched by very massive first stars exploding as pair-instability SNe (PISN). This star has indeed a very low abundance of Zn, a characteristic of the ejecta of PISN. 
This might be also possible for the matter that formed CS\,30302-145. We thus tried to measure the Zn abundance in CS\,30302-145 in order to detect  a Zn deficiency. We found  [Zn/Fe] $< +1$, this does not allow us to decide if the star is Zn poor or not. In extremely metal-poor stars, it is in fact impossible to detect a deficiency of Zn with an S/N as low as about 65 around the zinc line at 481\,nm. 

In CS\,30302-145, we were only able to measure the abundance of Sr of the neutron-capture elements.  It is highly deficient, but at this metallicity, many normal metal-poor stars are Sr-poor \citep[see e.g. ][]{FrancoisDH07}.  

We note that the star CS\,30302-145 has a retrograde orbit (see Fig. \ref{fig:orbit}\,A) and that its energy is relatively high (Fig. \ref{fig:orbit}\,E). As a consequence, this star may have been formed in a dwarf galaxy and have later been accreted by the Milky Way.

\section{Conclusions}

The aim of this study was to perform a detailed chemical analysis for a sample of 11 dwarf halo stars that extends from main sequence to turnoff, in order to detect abundance anomalies in Li-poor stars. 

A few of these very metal-poor stars ([Fe/H]$\leq$--3.0) are indeed characterised by a lower Li abundance than the Spite plateau.  We were able to compare the chemical pattern of stars on this plateau to the  Li-depleted stars, and we did not find any significant difference.

The star CS\,22882-027, in which the lithium line is not visible, seems to be a canonical blue straggler, as suggested by \citet{ryan01}, and \citet{bonifacio19}. 
In this star, the Sc abundance is rather low.

The CEMP star HE\,1413-1954 belongs to the low-carbon band, and as expected, it is a CEMP-no  \citep[without enrichment in heavy elements, see ][]{BonifacioCS15,bonifacio18}. We note that the Li abundance in this star is compatible with the Spite plateau \citep{MasseronJL12,MatsunoAS17}.

The star CS\,30302-145 shows low abundances of all the $\alpha$ elements except for Ca, but it has a high value of Mn. A similar behaviour was observed in the $\alpha$-poor EMP star ([Fe/H]=--4) HE\,1424-0241 \citet{cohen07}, but with some differences.\\
~~i/ The odd elements Na and Al are over-deficient in CS\,30302-145, but not in  HE\,1424-0241.\\
~ii/ Unlike in CS\,30302-145,  Mg is normal in HE\,1424-0241 and Ca is deficient.\\ 
iii/ The relatively high value of the energy of the orbit of CS\,30302-145 might indicate that this star has been accreted, but this is not the case of HE\,1424-0241.

The age of all the stars in our sample extends from 12 to 14 Gyr. 
(One star, CS\,22882-027, seems to be younger, but we have shown that it is very likely a blue straggler.)

In the study of the kinematic and orbital properties of the stars, we observed (Table \ref{tab:accret}) that the parameters of 6 of the 11 stars are compatible to an accretion by our Galaxy. The orbital parameters of only one star, the C-rich star HE\,1413-1954, suggest that this star was very probably formed in situ.\\
The blue straggler CS\,22882-027 and the $\alpha$-poor star CS\,30302-145 may have been formed in another galaxy and been accreted later.\\ 
Considered collectively, the Li-poor stars do not have any specific orbital characteristics. All have highly eccentric orbits, and they are almost equally divided between prograde and retrograde stars.


\begin{acknowledgements}
We gratefully acknowledge support from the French National Research Agency (ANR) funded project ``Pristine'' (ANR-18-CE31-0017).\\
This work has made use of data from the European Space Agency (ESA) mission
{\it Gaia} (\url{https://www.cosmos.esa.int/gaia}), processed by the {\it Gaia}
Data Processing and Analysis Consortium (DPAC,
\url{https://www.cosmos.esa.int/web/gaia/dpac/consortium}). Funding for the DPAC
has been provided by national institutions, in particular the institutions
participating in the {\it Gaia} Multilateral Agreement.
\end{acknowledgements}

\begin{appendix}
\section{Comparison with the recent literature}
\begin{itemize}

\item{BS\,17572-100 (or HE\,0926-0508):}
This star has been studied by \citet{barklem05} in the frame of the ESO-Hambourg survey, and \citet{melendez10} determined its Li abundance. \citet{zhang11} determined the abundance of Si in this star, taking the NLTE effects into account. When we consider that our value of A(Si) has to be increased by 0.25 for NLTE correction and that the log gf value adopted by \citet{zhang11} is 0.05 lower than our value (from Vald3), the agreement between our determinations and theirs is very good ($ \rm A(Si)_{NLTE}= 4.94~{\rm and}~ 5.01,$ rectively).\\

\item{CS\,22950-173:} This is  one of the blue metal-poor stars for which   \citet{preston00} searched for a variation in radial velocity. They concluded that it was not a spectroscopic binary. They found a metallicity [Fe/H] 0.2\,dex higher than our value, but this is explained by the fact that they adopted a temperature (based on the old relation $B-V$ color versus \Teff, following \citet{Boehm-Vitense81}) almost 500\,K higher than ours. \citet{spite14} also found a metallicity slightly higher than ours, but this is also due to an adopted higher temperature, which in this case was based on the profile of the $H_\alpha$ wings. Moreover, in both cases, the Fe lines list is not exactly the same.\\

\item{CS\,29514-007:} 
This star is considered a CEMP-no star in the extensive study of EMP stars carried out by \citet{roederer14}. We obtained the same metallicity as \citet{roederer14} for this star, but they found A(C)=6.49 while we were only able to estimate an upper limit of the carbon abundance:  ${\rm A(C)}\le 6.33$. We do not have a clear explanation for this difference in the A(C) determination. It might be that the G-band is so weak that the C determination is uncertain in any case. Unfortunately, unlike ESO, Magellan does not have a public archive, so we were unable to directly compare the UVES spectrum we analysed with that of Mike. In any case, this  star is not  a true CEMP star according to the  \citet{BeersChrist05} definition: metal-poor star with [C/Fe]> +1.\\

\item{HE\,0148-2611:}
This star was studied by  \citet{carretta02} from HIRES Keck spectra. With about the same atmospheric parameters \citep{cohen02} as those we deduced from the Gaia photometry, they obtained a lower metallicity [Fe/H]=--3.5 instead of [Fe/H]=--3.2 (Table \ref{tab:Abundance}).\\

\item{HE\,1413-1954:} 
This star is classified as a CEMP star following \citet{masseron10} and \citet{suda11}. Our values agree fairly well with those of \citet{masseron10}. Our higher value of the [C/Fe] ratio can be explained by the large uncertainty of the C abundance due to to the low S/N of the spectrum in the region of the CH band. This star was also analysed by  \citet{allen12}, and their abundances agree with our determinations.\\
The C abundance determination by \citet{zhang11} for this star perfectly agrees with our value. 
They also computed an NLTE profile of the Si 3905 line, however, and their A(Si) value was found to be much higher than ours. If we apply a NLTE correction \citep[+0.25 dex following ][]{ShiGM09} to our value, we find $\rm A(Si)_{NLTE}=4.25,$ to be compared to the value $\rm A(Si)_{NLTE}=5.04$, after correcting the log gf value by 0.05 (see BS\,17572-100 at the beginning of this section). These values are not compatible, although the same ESO-UVES spectra were used. The slight difference in the adopted effective temperature cannot explain this effect. In Fig \ref{fig:siHE1413}, we show the fit of our profile of the Si line.

\begin{figure}
\centering
\resizebox{8.1cm}{!}
{\includegraphics{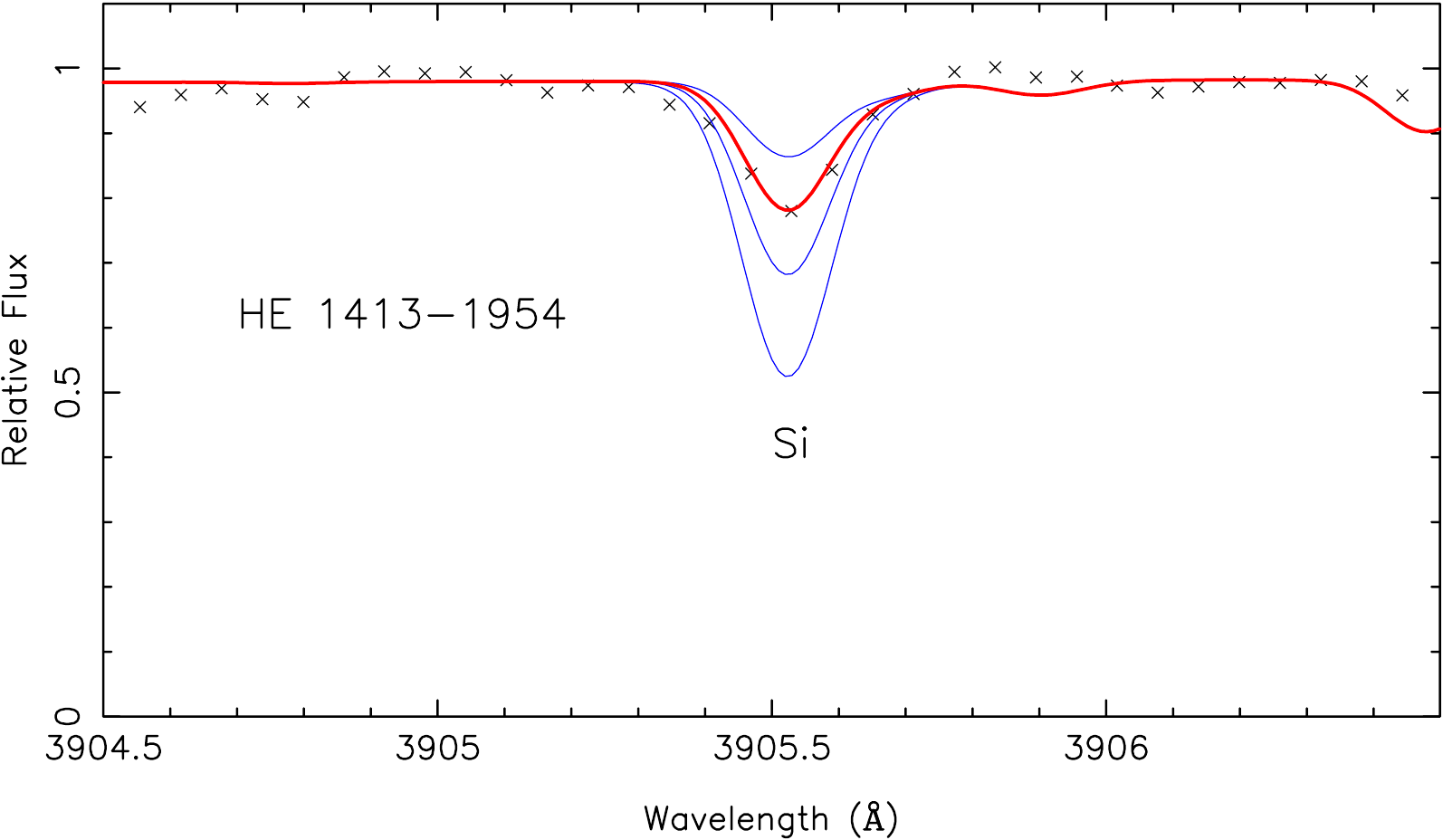}}
\caption{LTE profile of the Si line in HE\,1413-1954 computed with A(Si) = 3.7, 4.3, and 4.8 (blue lines) and with the best-fit value A(Si)=4.01 (red line). The NLTE value of A(Si)=5.09 would in this figure correspond to an LTE A(Si) value of about 4.8, which is incompatible with the observed profile,}
\label{fig:siHE1413}
\end{figure}

\end{itemize}
\end{appendix}

\end{document}